\documentclass[conference]{IEEEtran}
\IEEEoverridecommandlockouts
\usepackage{cite}
\usepackage{amsmath,amssymb,amsfonts}
\usepackage{algorithmic}
\usepackage{graphicx}
\usepackage{textcomp}
\usepackage[dvipsnames]{xcolor}
\usepackage{tabularx}
\usepackage{subfigure}
\usepackage{url}
\usepackage{booktabs}
\usepackage{makecell}
\usepackage{pifont} % For \ding symbols
\newcommand{\xmark}{\textcolor{red}{\ding{55}}} % Red X
\renewcommand{\checkmark}{\textcolor{ForestGreen}{\ding{51}}} % Green checkmark

\usepackage{multirow}

\usepackage{listings}
\usepackage[letterpaper, top=0.75in, bottom=1.1in, left=0.62in, right=0.64in]{geometry}

\lstdefinestyle{ieeeStyle}{
    language=Python,
    basicstyle=\ttfamily\scriptsize,  % Fonte compacta para coluna dupla
    numbers=left,                     % Números de linha à esquerda
    numberstyle=\tiny\color{gray},
    stepnumber=1,
    numbersep=5pt,
    keywordstyle=\color{blue},
    commentstyle=\color{green!50!black},
    stringstyle=\color{red!70!black},
    showstringspaces=false,
    breaklines=true,
    frame=none,                        % Remove moldura
    xleftmargin=2.5em,                  % Espaço para linha vertical
    framexleftmargin=2em,
    captionpos=b
}

\def\BibTeX{{\rm B\kern-.05em{\sc i\kern-.025em b}\kern-.08em
    T\kern-.1667em\lower.7ex\hbox{E}\kern-.125emX}}
\begin{document}

\title{Assessing the Impact of Post-Quantum Digital Signature Algorithms on Blockchains
\thanks{This study was partially funded by the Conselho Nacional de Desenvolvimento Científico e Tecnológico (CNPq) – Brazil, by the Brazilian National Research and Education Network (RNP) under the Iliada Project, and by the Fundação de Amparo à Pesquisa do Estado do Rio Grande do Sul (FAPERGS) through grant no. 22/2551-0000841-0 (INOVA-RS). In addition, Roben Lunardi is supported by FAPERGS and IFRS, and holds a postdoctoral fellowship from CAPES (PIPD/CAPES).}
}

\author{
\IEEEauthorblockN{Alison Gonçalves Schemitt
% \orcidlink{0009-0001-3278-0019}
}
\IEEEauthorblockA{
\textit{PUCRS}\\
Porto Alegre, Brazil \\
alison.schemitt@edu.pucrs.br}
\and
\IEEEauthorblockN{Henrique Fan da Silva
% \orcidlink{0009-0002-9033-2633}
}
\IEEEauthorblockA{\textit{UNIPAMPA}\\
Alegrete, Brazil \\
henriquefan.aluno@unipampa.edu.br }
\and
\IEEEauthorblockN{Roben Castagna Lunardi
% \orcidlink{0000-0002-8118-0802}
}
\IEEEauthorblockA{\textit{IFRS and PUCRS}\\
Porto Alegre, Brazil \\
roben.lunardi@zonanorte.ifrs.edu.br}
\and
\IEEEauthorblockN{Diego Kreutz
% \orcidlink{0000-0003-0830-0238}
}
\IEEEauthorblockA{\textit{UNIPAMPA}\\
Alegrete, Brazil \\
diegokreutz@unipampa.edu.br}
\and
\IEEEauthorblockN{Rodrigo Brandão Mansilha 
% \orcidlink{0000-0002-2083-653X}
}
\IEEEauthorblockA{
\textit{UNIPAMPA}\\
Alegrete, Brazil \\
 mansilha@unipampa.edu.br}
\and
\IEEEauthorblockN{Avelino Francisco Zorzo
% \orcidlink{0000-0002-0790-6759}
}
\IEEEauthorblockA{
\textit{PUCRS}\\
Porto Alegre, Brazil \\
avelino.zorzo@pucrs.br}
}

\maketitle

\begin{abstract}
The advent of quantum computing threatens the security of traditional encryption algorithms, motivating the development of post-quantum cryptography (PQC). In 2024, the National Institute of Standards and Technology (NIST) standardized several PQC algorithms, marking an important milestone in the transition toward quantum-resistant security.
Blockchain systems fundamentally rely on cryptographic primitives to guarantee data integrity and transaction authenticity. However, widely used algorithms such as ECDSA, employed in Bitcoin, Ethereum, and other networks, are vulnerable to quantum attacks. Although adopting PQC is essential for long-term security, its computational overhead in blockchain environments remains largely unexplored.
In this work, we propose a methodology for benchmarking both PQC and traditional cryptographic algorithms in blockchain contexts. We measure signature generation and verification times across diverse computational environments and simulate their impact at scale. Our evaluation focuses on PQC digital signature schemes (ML-DSA, Dilithium, Falcon, Mayo, SLH-DSA, SPHINCS+, and Cross) across security levels 1 to 5, comparing them to ECDSA, the current standard in Bitcoin and Ethereum.
Our results indicate that PQC algorithms introduce only minor performance overhead at security level 1, while in some scenarios they significantly outperform ECDSA at higher security levels. For instance, ML-DSA achieves a verification time of 0.14 ms on an ARM-based laptop at level 5, compared to 0.88 ms for ECDSA. We also provide an open-source implementation to ensure reproducibility and encourage further research.
\end{abstract}


\section{Introduction}\label{sec:introduction}

In recent years, research and development in quantum computing have advanced significantly, leveraging quantum-mechanical principles to tackle computationally intractable problems \cite{upama:2022}. Quantum computers can efficiently execute Shor's algorithm \cite{Shor:2019}, which poses a critical threat to cryptographic systems that rely on widely deployed algorithms such as RSA and ECDSA.

To address this challenge, the National Institute of Standards and Technology (NIST) initiated a process in 2016 to standardize post-quantum cryptography (PQC) algorithms. This effort has so far culminated in the approval of two digital signature schemes: FIPS 204, the Module-Lattice-Based Digital Signature Algorithm (ML-DSA) \cite{fips-204}, formerly known as Dilithium; and FIPS 205, the Stateless Hash-Based Digital Signature Algorithm (SLH-DSA) \cite{fips-205}, formerly SPHINCS+. In addition, the Falcon algorithm \cite{fouque2018falcon} has been approved and will be standardized as FIPS 206, the Fast-Fourier Transform over NTRU-Lattice-Based Digital Signature Algorithm (FN-DSA).

Beyond these standardized schemes, we also consider Cross \cite{baldi2025cross} and Mayo \cite{beullens2021mayo}, two candidate algorithms from the additional rounds of the NIST signature standardization process. Despite this progress, the performance and practicality of these algorithms across heterogeneous hardware platforms, software stacks, and application environments remain insufficiently explored.
 
Blockchains rely heavily on cryptography to ensure both integrity and authenticity. The ECDSA algorithm, widely adopted in networks such as Bitcoin and Ethereum, has been shown to be particularly vulnerable to quantum attacks \cite{Shor:2019, Yang:2024}. This vulnerability highlights the urgent need to migrate blockchain systems toward post-quantum cryptographic (PQC) alternatives. Although the adoption of PQC algorithms is essential for long-term security, their computational overhead in blockchain environments remains insufficiently explored \cite{PQCinEthereum-based}, especially when compared to evaluations conducted in other domains \cite{evalPQCSignatures, feasibilityPQCinV2V, KyberDillithiumComparison, classicalVsPQCinPi4-desktop-laptop, benchmarkPQCPi4}.

\begin{table*}[h!]
    \centering
    \caption{Comparison among this and related works.}
    \label{tab:related-work-combined}
    \footnotesize
    \newcolumntype{Y}{>{\centering\arraybackslash}X}
    \begin{tabularx}{\textwidth}{l|c|c|c|c|c|c|c|Y|>{\centering\arraybackslash}m{1.8cm}}
        \hline
        \hline 
         & \multicolumn{1}{c|}{\textbf{Traditional}} & \multicolumn{5}{c|}{\textbf{Post-Quantum}} & \multicolumn{2}{c|}{\textbf{CPU Architecture}} &  \\ 
        \textbf{Ref.} & \textbf{ECDSA} & \textbf{ML-DSA}  \textbf{(Dilithium)} & \textbf{Falcon} & \textbf{Mayo} & \textbf{SLH-DSA (SPHINCS+)} & \textbf{Cross} & \textbf{X64} & \textbf{ARM} & \textbf{Blockchain focus} \\
        \hline
        \makecell{\textbf{This}\\\textbf{work}} & \checkmark   & \checkmark  & \checkmark & \checkmark & \checkmark & \checkmark & \checkmark & \checkmark & \checkmark \\
        \hline
        \cite{PQCinEthereum-based} & \checkmark  & \checkmark  & \checkmark & \xmark & \checkmark & \xmark & \checkmark & \xmark & \checkmark \\
        \hline
        \cite{evalPQCSignatures} & \checkmark   & \checkmark  & \checkmark & \xmark & \checkmark & \xmark & \checkmark & \xmark & \xmark \\
        \hline
        \cite{feasibilityPQCinV2V} & \checkmark   & \checkmark  & \checkmark & \xmark & \checkmark & \xmark & \checkmark & \checkmark & \xmark \\
        \hline
        \cite{KyberDillithiumComparison} & \checkmark   & \checkmark  & \xmark & \xmark & \xmark & \xmark & \checkmark & \xmark & \xmark \\
        \hline
        \cite{classicalVsPQCinPi4-desktop-laptop} & \checkmark   & \checkmark  & \checkmark & \checkmark & \checkmark & \xmark & \checkmark & \checkmark & \xmark \\
        \hline
        \cite{benchmarkPQCPi4} & \xmark   & \checkmark  & \checkmark & \xmark & \checkmark & \xmark & \checkmark & \checkmark & \xmark \\
        \hline \hline 
    \end{tabularx}
\end{table*}

We aim to assess the performance impact on blockchain systems when migrating from traditional digital signature algorithms (e.g., ECDSA, RSA) to PQC alternatives (e.g., ML-DSA, SLH-DSA). This paper makes three main contributions:

\begin{itemize}
\item \textbf{Methodology:} We propose a clear and reproducible methodology that combines (i) algorithm benchmarking to evaluate isolated performance for three fundamental operations (key pair generation, signing, and signature verification), and (ii) large-scale simulation to assess the impact of these operations on blockchain systems. To ensure transparency and foster further research, we provide a public repository with all implementations and datasets.

\item \textbf{Empirical Evaluation:} We present an extensive empirical evaluation comprising more than 31,000 runs across three computational environments (two laptops and one desktop), covering two CPU architectures (x64 and ARM) and two operating systems (macOS and Linux). Our analysis includes seven algorithm families, comprising 62 implementations representing all NIST-defined security levels \cite{nistPQCproposals} (Levels 1--3 and 5, since no proposals exist for Level 4), providing a comprehensive performance characterization.

\item \textbf{Findings:} Our results demonstrate the practical feasibility of integrating PQC algorithms into blockchain systems. The evidence indicates that PQC can be adopted without prohibitive overhead, while also strengthening the long-term security of blockchain networks.
\end{itemize}

This paper is organized as follows. Section \ref{sec:related-works} discusses related work and identifies gaps our research addresses. Section \ref{sec:implementation} describes our methodology and experimental setup. Section \ref{sec:results} analyzes the findings, and Section \ref{sec:considerations} concludes with future research directions.

\section{Related Works}\label{sec:related-works}

The cryptography community has extensively discussed the standardization of Post-Quantum Cryptography (PQC) algorithms in recent years. Table \ref{tab:related-work-combined} summarizes representative works that evaluate these algorithms. The PQC algorithms studied in these papers were either finalists or selected as NIST standards. In addition, works such as \cite{PQCinEthereum-based, evalPQCSignatures, feasibilityPQCinV2V, KyberDillithiumComparison, classicalVsPQCinPi4-desktop-laptop} also provide comparisons between PQC algorithms and traditional digital signature algorithms. 

These studies primarily assess the performance of signature generation and verification. Among them, only \cite{PQCinEthereum-based} explicitly considers blockchain. This work analyzes real-time transaction payloads from a local blockchain and applies them to create and verify signatures using PQC algorithms and ECDSA. 

The studies in \cite{evalPQCSignatures, KyberDillithiumComparison} compare traditional algorithms (e.g., ECDSA) with PQC algorithms (e.g., ML-DSA) across different security levels. The work in \cite{KyberDillithiumComparison} further evaluates versions with and without hardware optimizations (AVX2). Both studies show that ECDSA achieves the best performance among traditional algorithms, while ML-DSA consistently performs best among PQC alternatives.  

The works in \cite{classicalVsPQCinPi4-desktop-laptop, benchmarkPQCPi4} examine PQC algorithms in the context of IoT devices and general-purpose computers, with \cite{classicalVsPQCinPi4-desktop-laptop} also including comparisons with traditional algorithms. Both conclude with recommendations on algorithm suitability depending on the target device type (desktop, laptop, or IoT). The study in \cite{feasibilityPQCinV2V} evaluates PQC algorithms in the context of smart vehicles (V2X) and concludes that, despite the good performance of some algorithms, none of the tested schemes fully meet the strict requirements of V2X systems.  

Overall, related work highlights that ML-DSA performs comparably to, and in some cases better than, traditional algorithms such as ECDSA, although with higher memory consumption \cite{KyberDillithiumComparison, PQCinEthereum-based}. At the same time, results show that in constrained environments such as IoT, PQC algorithms may not satisfy all system requirements, particularly with respect to key storage and resource limitations \cite{feasibilityPQCinV2V}.  

Our work differs by directly comparing standardized and finalist NIST PQC algorithms with ECDSA, which remains the dominant digital signature scheme in blockchain systems. Unlike prior studies, we specifically analyze the performance impact of migrating from traditional to PQC algorithms on block verification in blockchain environments.

\section{Proposed Evaluation Methodology}\label{sec:implementation}

This section presents our methodology, implementation details, and experimental setup.

\begin{table*}[!htp]
\centering
\footnotesize
\caption{Analyzed Algorithms. Among the evaluated algorithms, only two support security Level 2, and none are available for Level 4.}
 \renewcommand{\arraystretch}{1.2} 
  \setlength{\tabcolsep}{2pt} 
 
\renewcommand{\arraystretch}{1.3}
\begin{tabular} {m{1cm}|m{.5cm}|m{2.6cm}|m{3cm}|m{2cm}|m{3cm}|m{1.5cm}|m{3cm}}
\hline
\hline
& & & \multicolumn{5}{c}{\textbf{Variants (46)}} \\ 
\textbf{PQC?} &\# & \textbf{Algorithm (16)}  & \textbf{Level 1 (14)} & \textbf{Level 2 (2)} & \textbf{Level 3 (14)} & \textbf{Level 4 (0)} & \textbf{Level 5 (16)} \\
\hline
No & 1 & ECDSA & P-256 & — & P-384   &  — & P-521 \\ \hline
\multirow{15}{*}{Yes} 
& 2 & ML-DSA$^{*}$ & — & ML-DSA-44 & ML-DSA-65 & — &  ML-DSA-87 \\ \cline{2-8}
& 3 & Dilithium$^{*}$ & — & Dilithium2 & Dilithium3  & — & Dilithium5 \\ \cline{2-8}
& 4 &Falcon & Falcon-512 & — & —  & —& Falcon-1024 \\ \cline{2-8}
& 5 &Falcon-padded & Falcon-padded-512 & — & —  & —& Falcon-padded-1024 \\ \cline{2-8}
& 6 & Mayo & MAYO-2 & — & MAYO-3  & —& MAYO-5 \\ \cline{2-8}
& 7 & SPHINCS+-SHA-s & SHA2-128s-simple & — & SHA2-192s-simple  & —& SHA2-256s- simple \\ \cline{2-8}
& 8 & SPHINCS+-SHA-f & SHA2-128f-simple & — & SHA2-192f-simple  & —& SHA2-256f-simple \\ \cline{2-8}
& 9 & SPHINCS+-SHAKE-s & SHAKE-128s-simple & — & SHAKE-192s-simple & — & SHAKE-256s-simple \\ \cline{2-8}
& 10 & SPHINCS+-SHAKE-f & SHAKE-128f-simple & — & SHAKE-192f-simple & — & SHAKE-256f-simple \\ \cline{2-8}
& 11 & Cross-rsdp-small & Cross-rsdp-128-small & — & Cross-rsdp-192-small & — & Cross-rsdp-256-small \\ \cline{2-8}
& 12 & Cross-rsdpg-small & Cross-rsdpg-128-small & — & Cross-rsdpg-192-small & — & Cross-rsdpg-256-small \\ \cline{2-8}
& 13 & Cross-rsdp-balanced & Cross-rsdp-128-balanced & — & Cross-rsdp-192-balanced & — & Cross-rsdp-256-balanced \\ \cline{2-8}
& 14 & Cross-rsdpg-balanced & Cross-rsdpg-128-balanced & — & Cross-rsdpg-192-balanced & — & Cross-rsdpg-256-balanced \\ \cline{2-8}
& 15 & Cross-rsdp-fast & Cross-rsdp-128-fast & — & Cross-rsdp-192-fast & — & Cross-rsdp-256-fast \\ \cline{2-8}
& 16 & Cross-rsdpg-fast & Cross-rsdpg-128-fast & — & Cross-rsdpg-192-fast & — & Cross-rsdpg-256-fast \\ \hline
\hline
\multicolumn{7}{l}{$^{*}$Includes legacy (pre-standardization) implementations   and current NIST-standardized.} \\ 
\end{tabular}

\label{tab_algoritmos_disponiveis} 
\end{table*}

\subsection{Workflow}\label{sec:etapas-fluxo}
 
\begin{figure}[htb]
    \centering
    \includegraphics[width=\columnwidth]{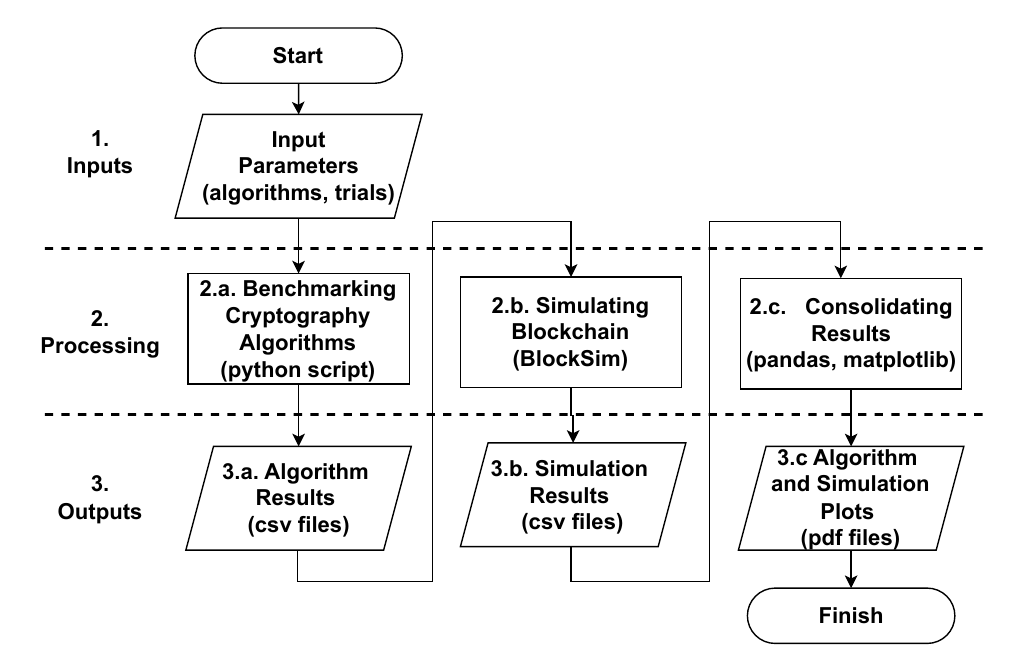}
    \caption{Methodological workflow.}
    \label{fig_etapas-fluxo}
\end{figure}

Figure~\ref{fig_etapas-fluxo} illustrates the methodological workflow, which is structured into three main components: (1) Inputs, (2) Processing, and (3) Outputs.  

The inputs consist of cryptographic algorithms, either traditional or post-quantum (PQC), along with their configurable parameters. We also define experimental parameters, such as the number of repetitions, to ensure statistically reliable results in both the algorithm benchmarking and blockchain simulation stages.  

The processing pipeline is organized into three sequential stages. In the first stage (2.a), focused on algorithm benchmarking, cryptographic algorithms are executed to measure key performance indicators, namely key generation, signing, and verification. These baseline measurements are then used in the second stage (2.b), the blockchain simulation, where the behavior of the algorithms is modeled during block verification in a distributed ledger environment. In the final stage (2.c), result synthesis, the outputs from both previous stages are consolidated and visualized to enable comparative analysis.  

The outputs consist of structured machine-readable datasets (e.g., CSV files) that contain both raw measurements and aggregated statistics, including mean and standard deviation. These datasets are complemented by graphical visualizations, such as bar charts, which support interpretation and highlight performance trade-offs across algorithms.

\subsection{Implementation}\label{sec:implementacao}

We present \texttt{PQCinBlock}\footnote{Source code available at: https://github.com/conseg/PQCinBlock}, a tool designed to evaluate both traditional and post-quantum cryptographic (PQC) algorithms. The tool is developed with two main objectives: (1) to enable comprehensive comparative analysis of cryptographic algorithms in both isolated and simulated blockchain environments, and (2) to provide extensibility for integrating new cryptographic schemes as they are developed.

\begin{figure}
    \centering
    \includegraphics[width=1\linewidth]{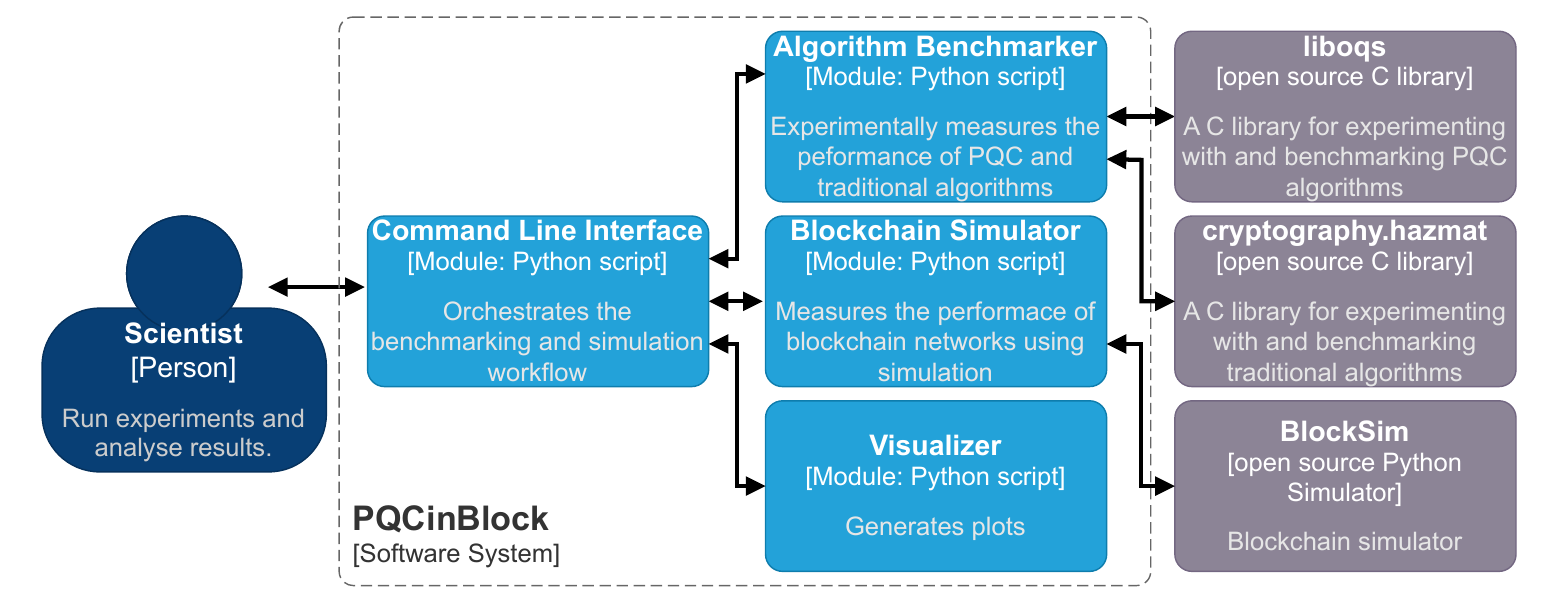}
    \caption{Architecture of the \texttt{PQCinBlock}.}
    \label{fig_blocksign}
\end{figure}

Figure~\ref{fig_blocksign} illustrates the architecture of \texttt{PQCinBlock}. The \texttt{Main} module receives and validates input parameters and orchestrates the other modules, which operate as described below.

The \texttt{Algorithm Benchmarker} module executes the selected algorithms in \textit{Python}, computing means and standard deviations for key generation, signature creation, and signature verification. It enables the integration of new algorithms and programming languages without requiring modifications to other modules. To add a new algorithm, users create a Python file in the \texttt{algorithms} directory describing the cryptographic variants, their security levels, and the functions to measure operational performance. Currently, we use the \texttt{liboqs} version 0.12.0 library for post-quantum algorithms and \texttt{cryptography} version 45.0.2 for the traditional ECDSA algorithm. Both libraries are wrappers for \textit{C/C++} implementations, enabling direct comparisons between traditional and post-quantum approaches.

The module currently supports the algorithms listed in Table \ref{tab_algoritmos_disponiveis}, including \texttt{ECDSA}, \texttt{ML-DSA}, several variants of \texttt{SPHINCS+} (\textit{e.g.}, \texttt{SPHINCS+-SHA-s}, \texttt{SPHINCS+-SHAKE-f}), \texttt{Falcon}, \texttt{Mayo}, and multiple implementations of \texttt{Cross-rsdp} and \texttt{Cross-rsdpg}. It is important to note that \texttt{ML-DSA} is equivalent to \texttt{Dilithium} under the nomenclature adopted by NIST in 2024 \cite{fips-todos}, while \texttt{Falcon-padded} is a variant of \texttt{Falcon} with fixed-length signatures.

The \texttt{Blockchain Simulator} models the behavior of blockchain networks. It leverages performance metrics from the \texttt{Algorithm Benchmarker} module to assess the impact of cryptographic algorithms on block verification processes. Currently, we employ \texttt{BlockSim} \cite{blocksim}, an open-source blockchain simulation framework that captures the behavior of key blockchain layers, including network propagation, consensus mechanisms, and incentive structures. The \texttt{PQCinBlock} framework, using the \texttt{BlockSim} simulator, currently supports two blockchain models: Bitcoin and Ethereum.

The \texttt{Visualizer} module processes raw output data from the other components, generating plots that enable comparative analysis of algorithm performance across different evaluation scenarios. It currently produces bar plots to display both benchmarking results and blockchain simulation outcomes.

\subsection{Instantiation}\label{sec:methodology}

The experiments were conducted across multiple hardware configurations (detailed in Table~\ref{tab:machines}) to minimize potential biases introduced by specific CPU architectures or operating systems. 

\begin{table}
    \centering
    \scriptsize
    \caption{System environments.}
    \setlength{\tabcolsep}{2pt} 
    \renewcommand{\arraystretch}{1.1}
    \begin{tabularx}{\columnwidth}{>{\centering\arraybackslash}m{1cm}|
    >{\centering\arraybackslash}m{2cm}|
    c|
    >{\centering\arraybackslash}m{2.8cm}|
    >{\centering\arraybackslash}m{1.5cm}}
        \hline
        \hline
        \textbf{Machine} & \textbf{CPU} & \textbf{Mem.} & \textbf{OS} & \textbf{Timer Resolution}  \\
        \hline
        Laptop ARM & Apple M1 & 8 GB & macOS Darwin Kernel 24.0.0 & 40.978193 ns \\
        \hline
        Laptop x64 & Intel Core i7-1360P & 32 GB & Ubuntu  22.04.1 LTS Linux Kernel 6.8.0-65-generic & 37.980498 ns \\
        \hline
            Desktop  & AMD Ryzen 7 5800X & 80 GB & Ubuntu 24.04.2 LTS Linux Kernel 6.8.0-64-generic & 49.971276 ns \\
        \hline
        \hline
    \end{tabularx}
    \label{tab:machines}
\end{table}

Our experimental methodology included comprehensive benchmarking of all available algorithm variants, including nominally distinct but functionally equivalent implementations. The evaluation framework simulated Bitcoin and an Ethereum-based blockchain environment. To ensure statistical validity, we executed 1,000 warm-up iterations (which were discarded), as well as 10,000 measured executions per algorithm variant and 1,000 simulations. Parameters are listed in Table~\ref{tab_argumentos}. 

\begin{table}[!htp]
\centering
\scriptsize
\caption{Evaluation parameters.}
\label{tab_argumentos} 
  \setlength{\tabcolsep}{2pt} 
\renewcommand{\arraystretch}{1.2}
\begin{tabular} {m{2.4cm}|m{0.9cm}|m{4.3cm}}
\hline
\hline
\textbf{Argument} & \textbf{Value(s)} & \textbf{Description}\\
\hline
% \texttt{--help}, \texttt{-h}, & -- &  Displays the help message with the description of all available arguments and instructions on how to use the tool. \\ \hline
% \texttt{--list-sign} & -- &  Displays all signature algorithms available in the tool. \\ \hline
% \texttt{--sign} & -- &  List of digital signature algorithms to be evaluated. Supports multiple values, including traditional (i.e., ECDSA) and post-quantum (e.g., Dilithium, Falcon, SPHINCS+) algorithms. \\ \hline
\texttt{--runs}, \texttt{-r} & 10,000 &  Number of executions of each algorithm.  \\ \hline
\texttt{--warm-up}, \texttt{-wp} & 1,000 & Number of warm-up runs before the main measurement, for performance stabilization. \\ \hline
\texttt{--levels}, \texttt{-l} & \{1, 2, 3, 5\} & Defines the NIST security levels (1 to 5) of the algorithms to be tested. Can receive multiple values. \\ \hline
\texttt{--runs-simulator} & 1,000 & Number of simulation runs in BlockSim. \\ \hline 
\texttt{--model} & \{1, 2\} & Bitcoin, Etherium networks. \\
\hline
\hline
\end{tabular}
\end{table}

\section{Evaluation Results}\label{sec:results}

We present our results through three sequential components: (\textit{i}) benchmarking of cryptographic algorithms, (\textit{ii}) system-level blockchain simulations, and (\textit{iii}) interpretation of findings, including the identification of methodological constraints.

\begin{table*}[htb]
    \centering
    \scriptsize
    \caption{Complete Results on the Performance of Cryptography Algorithms (ms).}
    \label{tab:benchmark}
    \renewcommand{\arraystretch}{1.1}
    \setlength{\tabcolsep}{1pt} 
    \resizebox{\textwidth}{!}{
    \begin{tabular}{l|ccc|ccc|ccc}
    \hline 
    \hline 
         &  \multicolumn{3}{c|}{Lower Level (1 or 2)} & \multicolumn{3}{c|}{Level 3} & \multicolumn{3}{c}{Level 5}   \\ 
Algorithm & keypair (mean ± std) & sign (mean ± std) & verify (mean ± std) & keypair (mean ± std) & sign (mean ± std) & verify (mean ± std) & keypair (mean ± std) & sign (mean ± std) & verify (mean ± std) \\  

\hline \hline \multicolumn{10}{c}{Laptop ARM} \\  \hline 
ECDSA & 0.0198 ± 0.001 & 0.0441 ± 0.0021 & 0.0788 ± 0.0029 & 0.2426 ± 0.0091 & 0.3758 ± 0.0122 & 0.7366 ± 0.0162 & 0.2838 ± 0.0089 & 0.4973 ± 0.0157 & 0.8847 ± 0.0202 \\ 
ML-DSA & 0.049 ± 0.0041 & 0.2113 ± 0.1409 & 0.0529 ± 0.0026 & 0.0934 ± 0.0033 & 0.3425 ± 0.2338 & 0.0847 ± 0.0045 & 0.1351 ± 0.0056 & 0.4295 ± 0.2668 & 0.1402 ± 0.0055 \\ 
Dilithium & 0.0218 ± 0.0013 & 0.0653 ± 0.0407 & 0.0221 ± 0.0013 & 0.0465 ± 0.0026 & 0.0979 ± 0.0604 & 0.0329 ± 0.0021 & 0.0537 ± 0.0026 & 0.1195 ± 0.061 & 0.0522 ± 0.0021 \\ 
Falcon & 5.2091 ± 1.2291 & 0.155 ± 0.0043 & 0.029 ± 0.0036 &  &  &  & 16.7976 ± 4.2397 & 0.3071 ± 0.0062 & 0.0557 ± 0.0044 \\ 
Falcon-padded & 5.1996 ± 1.1972 & 0.1552 ± 0.0044 & 0.0286 ± 0.0022 &  &  &  & 16.7562 ± 4.226 & 0.3082 ± 0.0102 & 0.0551 ± 0.0029 \\ 
Mayo & 0.2576 ± 0.0033 & 0.3999 ± 0.0046 & 0.0342 ± 0.0015 & 0.3602 ± 0.0051 & 1.0552 ± 0.0125 & 0.173 ± 0.0028 & 0.9739 ± 0.0082 & 2.7203 ± 0.0161 & 0.379 ± 0.0045 \\ 
SPHINCS-SHA-s & 39.8194 ± 4.4966 & 302.4854 ± 1.319 & 0.3056 ± 0.0144 & 59.0928 ± 0.2914 & 581.6236 ± 2.6707 & 0.5038 ± 0.0143 & 39.0708 ± 0.2259 & 528.3311 ± 2.7981 & 0.7282 ± 0.0197 \\ 
SPHINCS-SHA-f & 0.6223 ± 0.0089 & 14.4995 ± 0.1365 & 0.8713 ± 0.0231 & 0.928 ± 0.0102 & 25.3311 ± 0.178 & 1.3657 ± 0.0246 & 2.4538 ± 0.023 & 52.2223 ± 0.3399 & 1.4073 ± 0.0273 \\ 
SPHINCS-SHAKE-s & 73.1857 ± 0.4665 & 556.0264 ± 6.5939 & 0.5522 ± 0.0245 & 108.2966 ± 0.8138 & 975.1893 ± 6.5981 & 0.8076 ± 0.0241 & 72.2639 ± 0.5518 & 860.007 ± 5.8928 & 1.1941 ± 0.0326 \\ 
SPHINCS-SHAKE-f & 1.1397 ± 0.0151 & 26.6009 ± 0.2408 & 1.5859 ± 0.0407 & 1.7003 ± 0.0197 & 43.8261 ± 0.3477 & 2.3626 ± 0.0443 & 4.5113 ± 0.0496 & 90.6262 ± 0.7514 & 2.4171 ± 0.0523 \\ 
Cross-rsdp-small & 0.0144 ± 0.001 & 3.0152 ± 0.0302 & 1.7568 ± 0.0198 & 0.0306 ± 0.0014 & 4.268 ± 0.0274 & 2.3046 ± 0.0197 & 0.052 ± 0.0017 & 6.1769 ± 0.0642 & 3.0694 ± 0.0358 \\ 
Cross-rsdpg-small & 0.0075 ± 0.0011 & 2.171 ± 0.0271 & 1.1833 ± 0.0156 & 0.0137 ± 0.0011 & 2.7817 ± 0.0207 & 1.7868 ± 0.0157 & 0.0219 ± 0.0012 & 3.7108 ± 0.0408 & 2.2536 ± 0.0261 \\ 
Cross-rsdp-balanced & 0.0141 ± 0.0009 & 0.7992 ± 0.0076 & 0.4615 ± 0.0052 & 0.0306 ± 0.0013 & 1.8272 ± 0.0149 & 1.0017 ± 0.0093 & 0.0527 ± 0.0018 & 3.3201 ± 0.0297 & 1.7039 ± 0.0168 \\ 
Cross-rsdpg-balanced & 0.0073 ± 0.0006 & 0.6097 ± 0.0059 & 0.3368 ± 0.0044 & 0.0137 ± 0.0008 & 0.7664 ± 0.0071 & 0.4722 ± 0.0054 & 0.0216 ± 0.0012 & 1.3359 ± 0.0111 & 0.8144 ± 0.0074 \\ 
Cross-rsdp-fast & 0.0141 ± 0.0007 & 0.4544 ± 0.0131 & 0.2603 ± 0.007 & 0.0306 ± 0.0013 & 1.0475 ± 0.0115 & 0.6051 ± 0.0086 & 0.0517 ± 0.0014 & 2.0195 ± 0.0171 & 1.1543 ± 0.0095 \\ 
Cross-rsdpg-fast & 0.0073 ± 0.0007 & 0.3221 ± 0.0034 & 0.1877 ± 0.0028 & 0.0136 ± 0.0008 & 0.6023 ± 0.0066 & 0.3743 ± 0.006 & 0.0221 ± 0.0009 & 1.1189 ± 0.0105 & 0.6675 ± 0.0068 \\ 
\hline \hline \multicolumn{10}{c}{Laptop x64} \\  \hline 
ECDSA & 0.012 ± 0.0009 & 0.0259 ± 0.001 & 0.0574 ± 0.0014 & 0.0775 ± 0.0013 & 0.1263 ± 0.0015 & 0.2512 ± 0.0024 & 0.0866 ± 0.0025 & 0.1754 ± 0.0046 & 0.3126 ± 0.0081 \\ 
ML-DSA & 0.0193 ± 0.0009 & 0.0493 ± 0.0277 & 0.0196 ± 0.0007 & 0.0322 ± 0.0011 & 0.0795 ± 0.0458 & 0.0315 ± 0.001 & 0.0491 ± 0.0017 & 0.0956 ± 0.0445 & 0.0479 ± 0.0016 \\ 
Dilithium & 0.0192 ± 0.0009 & 0.0488 ± 0.0277 & 0.0193 ± 0.0009 & 0.0322 ± 0.0014 & 0.0796 ± 0.0468 & 0.0315 ± 0.0011 & 0.0504 ± 0.0015 & 0.0977 ± 0.0452 & 0.0493 ± 0.0014 \\ 
Falcon & 5.8027 ± 1.5008 & 0.2039 ± 0.0034 & 0.0434 ± 0.0026 &  &  &  & 13.2128 ± 3.3529 & 0.2944 ± 0.0059 & 0.0604 ± 0.003 \\ 
Falcon-padded & 4.3956 ± 1.083 & 0.1555 ± 0.0058 & 0.0319 ± 0.0007 &  &  &  & 13.1972 ± 3.311 & 0.3002 ± 0.009 & 0.0592 ± 0.001 \\ 
Mayo & 0.0289 ± 0.001 & 0.063 ± 0.009 & 0.0142 ± 0.0022 & 0.0528 ± 0.0008 & 0.1867 ± 0.0023 & 0.0673 ± 0.0009 & 0.1188 ± 0.0017 & 0.36 ± 0.0046 & 0.1245 ± 0.0017 \\ 
SPHINCS-SHA-s & 17.4697 ± 0.1948 & 132.7452 ± 1.3186 & 0.1787 ± 0.0059 & 26.2132 ± 0.2847 & 253.8474 ± 1.932 & 0.298 ± 0.0069 & 17.4807 ± 0.1946 & 231.1655 ± 1.8035 & 0.4163 ± 0.0094 \\ 
SPHINCS-SHA-f & 0.2817 ± 0.0036 & 6.537 ± 0.0702 & 0.493 ± 0.0122 & 0.4189 ± 0.0048 & 11.3276 ± 0.1266 & 0.7512 ± 0.0132 & 1.1024 ± 0.0127 & 23.2572 ± 0.2587 & 0.7621 ± 0.0141 \\ 
SPHINCS-SHAKE-s & 36.7059 ± 0.1882 & 279.7878 ± 0.876 & 0.3461 ± 0.0108 & 53.8654 ± 0.4963 & 483.193 ± 1.5473 & 0.4903 ± 0.0122 & 35.6218 ± 0.1637 & 423.1682 ± 1.1481 & 0.7043 ± 0.0154 \\ 
SPHINCS-SHAKE-f & 0.577 ± 0.0062 & 13.4564 ± 0.1352 & 0.9368 ± 0.0208 & 0.8483 ± 0.0059 & 21.8742 ± 0.1583 & 1.3467 ± 0.0206 & 2.2406 ± 0.0109 & 44.9622 ± 0.2045 & 1.3681 ± 0.022 \\ 
Cross-rsdp-small & 0.0155 ± 0.0007 & 2.0538 ± 0.0168 & 1.5309 ± 0.014 & 0.0295 ± 0.0008 & 2.955 ± 0.0274 & 2.1288 ± 0.0233 & 0.0439 ± 0.0013 & 3.9732 ± 0.0433 & 2.839 ± 0.0353 \\ 
Cross-rsdpg-small & 0.0086 ± 0.0006 & 1.425 ± 0.014 & 0.911 ± 0.0133 & 0.0146 ± 0.0008 & 2.3393 ± 0.0275 & 1.4576 ± 0.017 & 0.0222 ± 0.0008 & 2.9133 ± 0.0367 & 1.7825 ± 0.0241 \\ 
Cross-rsdp-balanced & 0.0145 ± 0.0011 & 0.5602 ± 0.0104 & 0.3905 ± 0.0071 & 0.03 ± 0.0043 & 1.2859 ± 0.0166 & 0.8724 ± 0.0093 & 0.0463 ± 0.0016 & 2.148 ± 0.026 & 1.4411 ± 0.018 \\ 
Cross-rsdpg-balanced & 0.0082 ± 0.001 & 0.4135 ± 0.0069 & 0.2599 ± 0.0024 & 0.0141 ± 0.0006 & 0.6513 ± 0.0095 & 0.4012 ± 0.0094 & 0.0216 ± 0.0008 & 1.0667 ± 0.0158 & 0.6573 ± 0.0088 \\ 
Cross-rsdp-fast & 0.0137 ± 0.0006 & 0.337 ± 0.0084 & 0.2101 ± 0.0023 & 0.0296 ± 0.0007 & 0.7728 ± 0.0087 & 0.4858 ± 0.0057 & 0.0458 ± 0.001 & 1.3681 ± 0.012 & 0.8397 ± 0.0076 \\ 
Cross-rsdpg-fast & 0.0081 ± 0.0006 & 0.2336 ± 0.0138 & 0.1457 ± 0.0095 & 0.0137 ± 0.0006 & 0.5491 ± 0.0062 & 0.3446 ± 0.0039 & 0.0212 ± 0.0007 & 0.8832 ± 0.0129 & 0.5626 ± 0.0088 \\ 
\hline \hline \multicolumn{10}{c}{Desktop} \\  \hline 
ECDSA & 0.015 ± 0.001 & 0.033 ± 0.0016 & 0.0615 ± 0.0019 & 0.1218 ± 0.0045 & 0.1992 ± 0.0067 & 0.3977 ± 0.011 & 0.1087 ± 0.0027 & 0.2243 ± 0.0042 & 0.3909 ± 0.0067 \\ 
ML-DSA & 0.0236 ± 0.0008 & 0.0553 ± 0.0297 & 0.0226 ± 0.0006 & 0.0387 ± 0.0017 & 0.0873 ± 0.0473 & 0.0368 ± 0.0014 & 0.0605 ± 0.0019 & 0.1071 ± 0.0455 & 0.0578 ± 0.0017 \\ 
Dilithium & 0.0234 ± 0.0008 & 0.0548 ± 0.0301 & 0.0225 ± 0.0006 & 0.0386 ± 0.0008 & 0.086 ± 0.0459 & 0.0368 ± 0.0007 & 0.0605 ± 0.0011 & 0.1086 ± 0.0467 & 0.0579 ± 0.001 \\ 
Falcon & 4.5585 ± 1.1817 & 0.1584 ± 0.0022 & 0.034 ± 0.002 &  &  &  & 13.5104 ± 3.46 & 0.3151 ± 0.0035 & 0.0653 ± 0.0029 \\ 
Falcon-padded & 4.5356 ± 1.1917 & 0.1619 ± 0.0021 & 0.0338 ± 0.0025 &  &  &  & 13.4322 ± 3.3585 & 0.3197 ± 0.0078 & 0.0645 ± 0.001 \\ 
Mayo & 0.0237 ± 0.0007 & 0.0582 ± 0.001 & 0.0133 ± 0.0004 & 0.0418 ± 0.0013 & 0.1542 ± 0.0015 & 0.0473 ± 0.001 & 0.0884 ± 0.002 & 0.3045 ± 0.0031 & 0.0921 ± 0.0012 \\ 
SPHINCS-SHA-s & 15.1549 ± 0.0525 & 115.2267 ± 0.2323 & 0.166 ± 0.0056 & 22.1562 ± 0.1061 & 212.136 ± 0.8036 & 0.278 ± 0.0066 & 14.8261 ± 0.1126 & 192.8551 ± 1.4157 & 0.3829 ± 0.0077 \\ 
SPHINCS-SHA-f & 0.2407 ± 0.0033 & 5.5964 ± 0.021 & 0.4355 ± 0.0083 & 0.3587 ± 0.0033 & 9.5745 ± 0.0147 & 0.6604 ± 0.0083 & 0.9285 ± 0.0034 & 19.4653 ± 0.0334 & 0.6719 ± 0.01 \\ 
SPHINCS-SHAKE-s & 32.1903 ± 0.2419 & 245.0352 ± 1.7179 & 0.3163 ± 0.0108 & 49.5265 ± 0.3534 & 444.061 ± 3.1006 & 0.4499 ± 0.0113 & 29.7797 ± 0.0936 & 359.8339 ± 0.6492 & 0.6277 ± 0.0143 \\ 
SPHINCS-SHAKE-f & 0.4978 ± 0.0061 & 11.6084 ± 0.0367 & 0.8352 ± 0.0213 & 0.7429 ± 0.0038 & 19.1604 ± 0.052 & 1.228 ± 0.0171 & 1.887 ± 0.0123 & 38.123 ± 0.0904 & 1.1994 ± 0.0206 \\ 
Cross-rsdp-small & 0.0154 ± 0.0007 & 1.9686 ± 0.0071 & 1.4158 ± 0.0058 & 0.0313 ± 0.0008 & 2.7865 ± 0.0118 & 1.9685 ± 0.0072 & 0.0419 ± 0.0011 & 3.7086 ± 0.0116 & 2.583 ± 0.0094 \\ 
Cross-rsdpg-small & 0.0091 ± 0.0006 & 1.3865 ± 0.0054 & 0.8851 ± 0.0041 & 0.015 ± 0.0008 & 2.243 ± 0.0101 & 1.4572 ± 0.0064 & 0.0216 ± 0.0007 & 2.7278 ± 0.0091 & 1.6763 ± 0.005 \\ 
Cross-rsdp-balanced & 0.0151 ± 0.0009 & 0.5354 ± 0.0036 & 0.3674 ± 0.0028 & 0.031 ± 0.001 & 1.1966 ± 0.0044 & 0.8251 ± 0.0033 & 0.0417 ± 0.0012 & 1.9548 ± 0.0068 & 1.3485 ± 0.0052 \\ 
Cross-rsdpg-balanced & 0.0086 ± 0.0006 & 0.3987 ± 0.0023 & 0.248 ± 0.0018 & 0.0147 ± 0.0007 & 0.6246 ± 0.0025 & 0.385 ± 0.0019 & 0.0208 ± 0.0007 & 1.0061 ± 0.0051 & 0.6172 ± 0.0039 \\ 
Cross-rsdp-fast & 0.0156 ± 0.0006 & 0.3178 ± 0.0021 & 0.1968 ± 0.0017 & 0.03 ± 0.0008 & 0.7184 ± 0.0043 & 0.4553 ± 0.0027 & 0.043 ± 0.0009 & 1.2472 ± 0.0066 & 0.794 ± 0.0037 \\ 
Cross-rsdpg-fast & 0.0084 ± 0.0005 & 0.2208 ± 0.0017 & 0.1359 ± 0.0013 & 0.0145 ± 0.0007 & 0.5276 ± 0.0048 & 0.3249 ± 0.003 & 0.0205 ± 0.0007 & 0.8221 ± 0.0026 & 0.5191 ± 0.0022 \\ 

\hline 
\hline 
\end{tabular}}
\end{table*}

\subsection{Benchmark of Cryptographic Algorithms}

Table~\ref{tab:benchmark} presents the complete benchmark results, including the mean execution time and standard deviation for each algorithm implementation across all available security levels. These metrics were calculated over the specified number of runs for key-pair generation, signing, and verification operations. Due to space constraints, ML-DSA Level 2 results are shown in the column designated for lower levels (1 or 2), as ML-DSA does not provide a Level 1 variant.

The comprehensive results in Table~\ref{tab:benchmark} illustrate the complexity of the study, while cross-machine consistency confirms the reliability of our methodology. Among the higher-cost PQC algorithms, the top performers (ML-DSA and Mayo) were selected for further analysis.

\begin{figure*}[htb]
\centering
\subfigure[Laptop ARM - Lower Level (1 or 2)]{
\includegraphics[width=.31\textwidth]{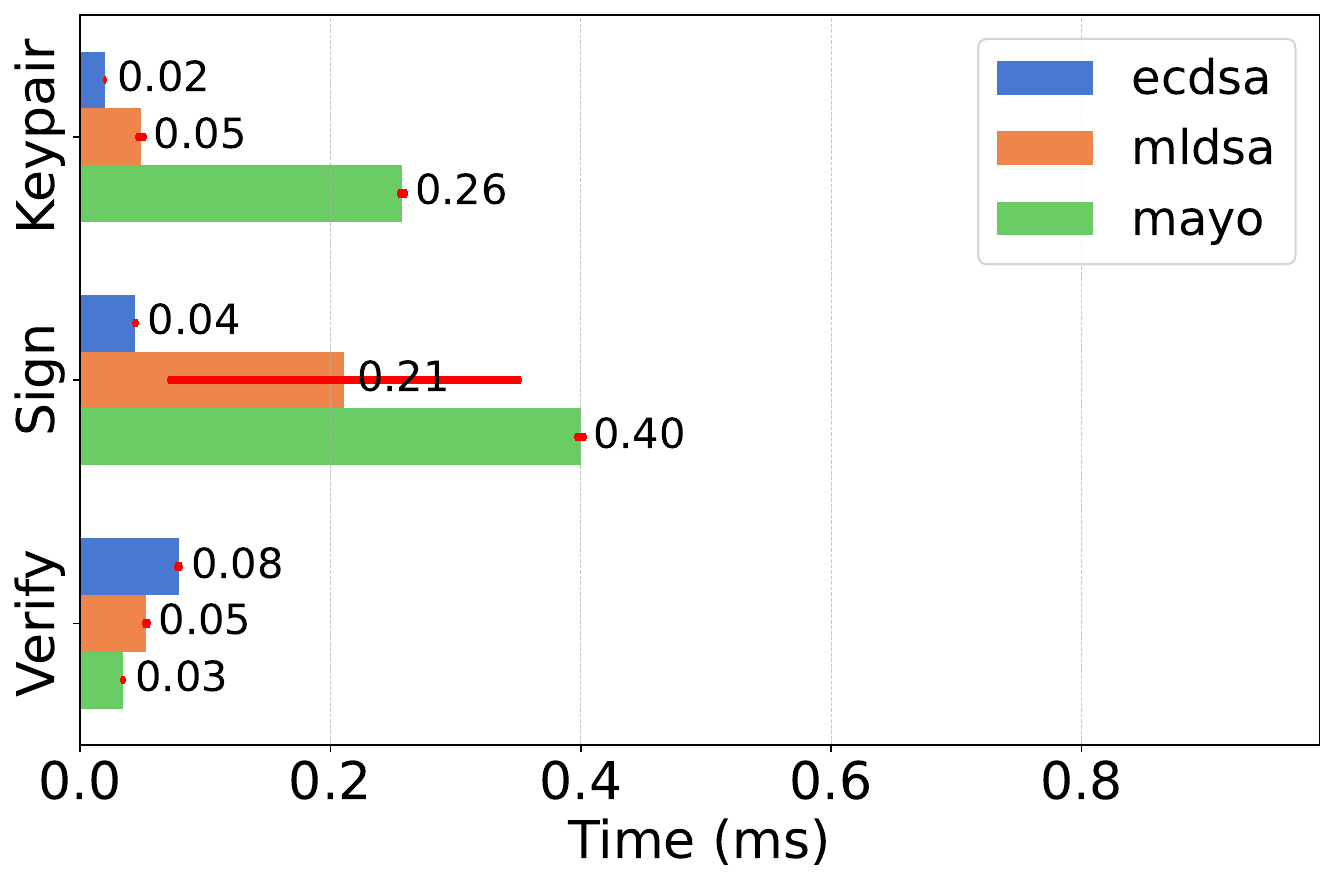}
\label{fig_m1-execucoes-level-1}
}
\subfigure[Laptop ARM - Level 3]{
\includegraphics[width=.31\textwidth]{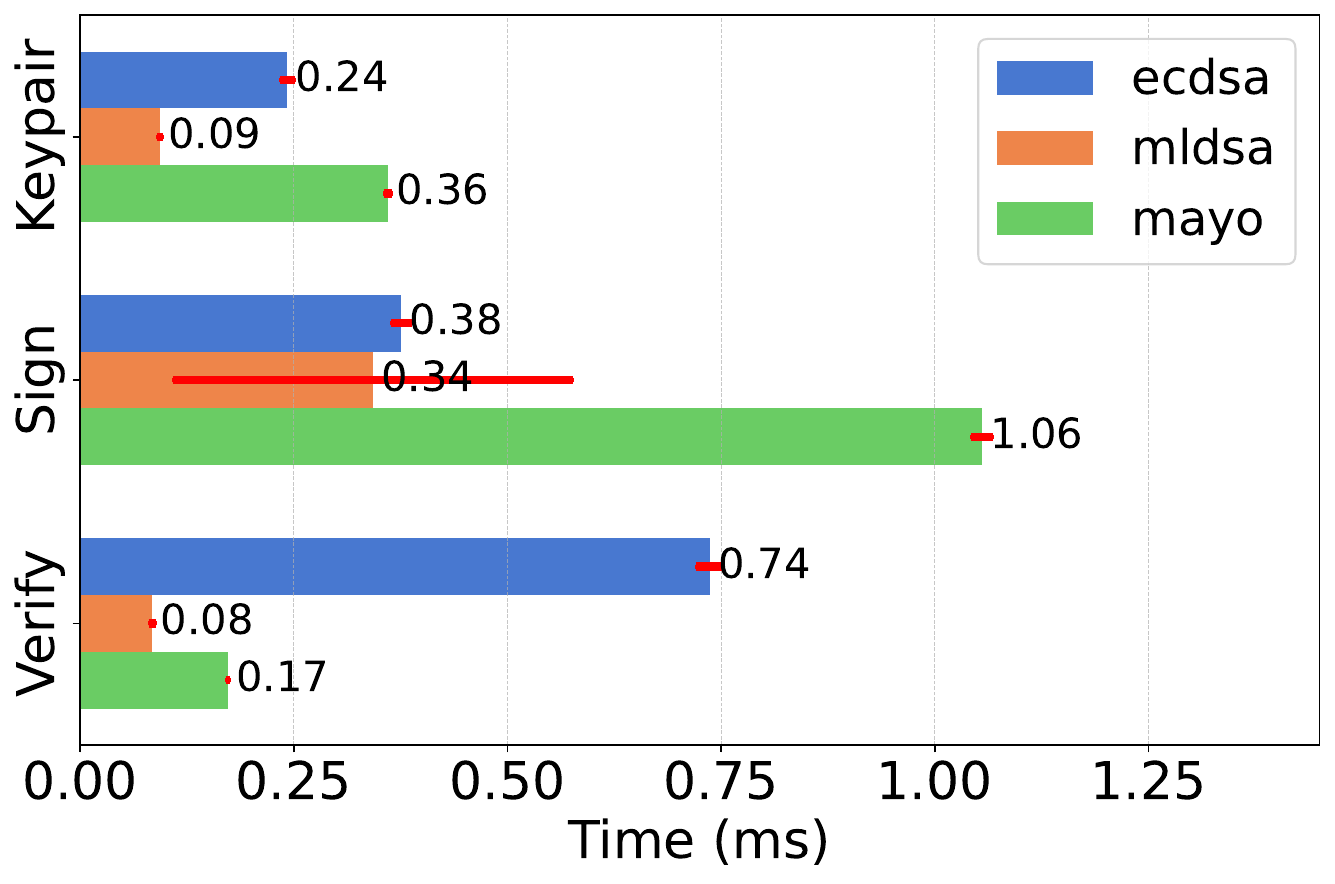}
\label{fig_m1-execucoes-level-3}
}
\subfigure[Laptop ARM - Level 5]{
\includegraphics[width=.31\textwidth]{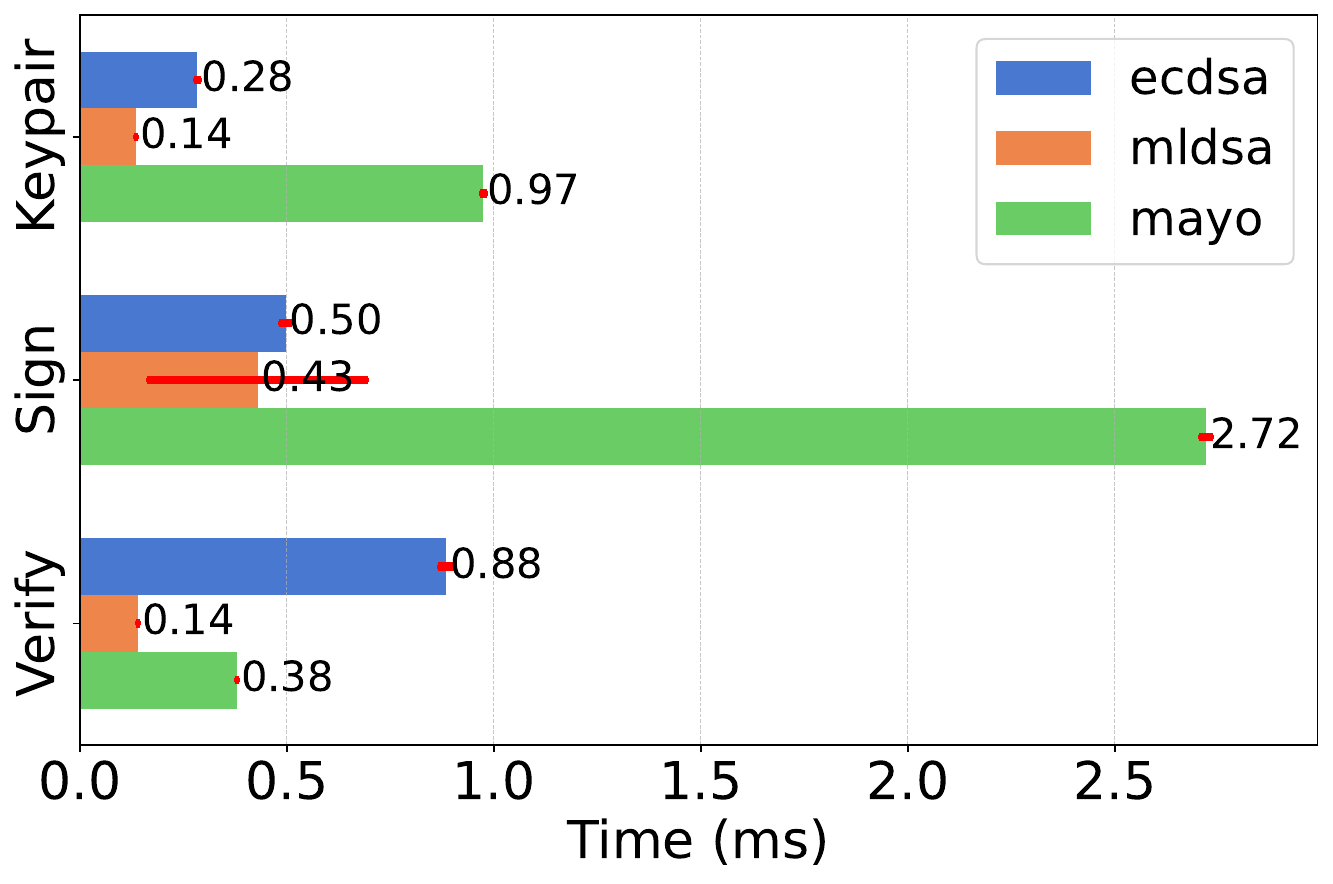}
\label{fig_m1-execucoes-level-5}
}
\subfigure[Laptop x64 - Lower Level (1 or 2)]{
\includegraphics[width=.31\textwidth]{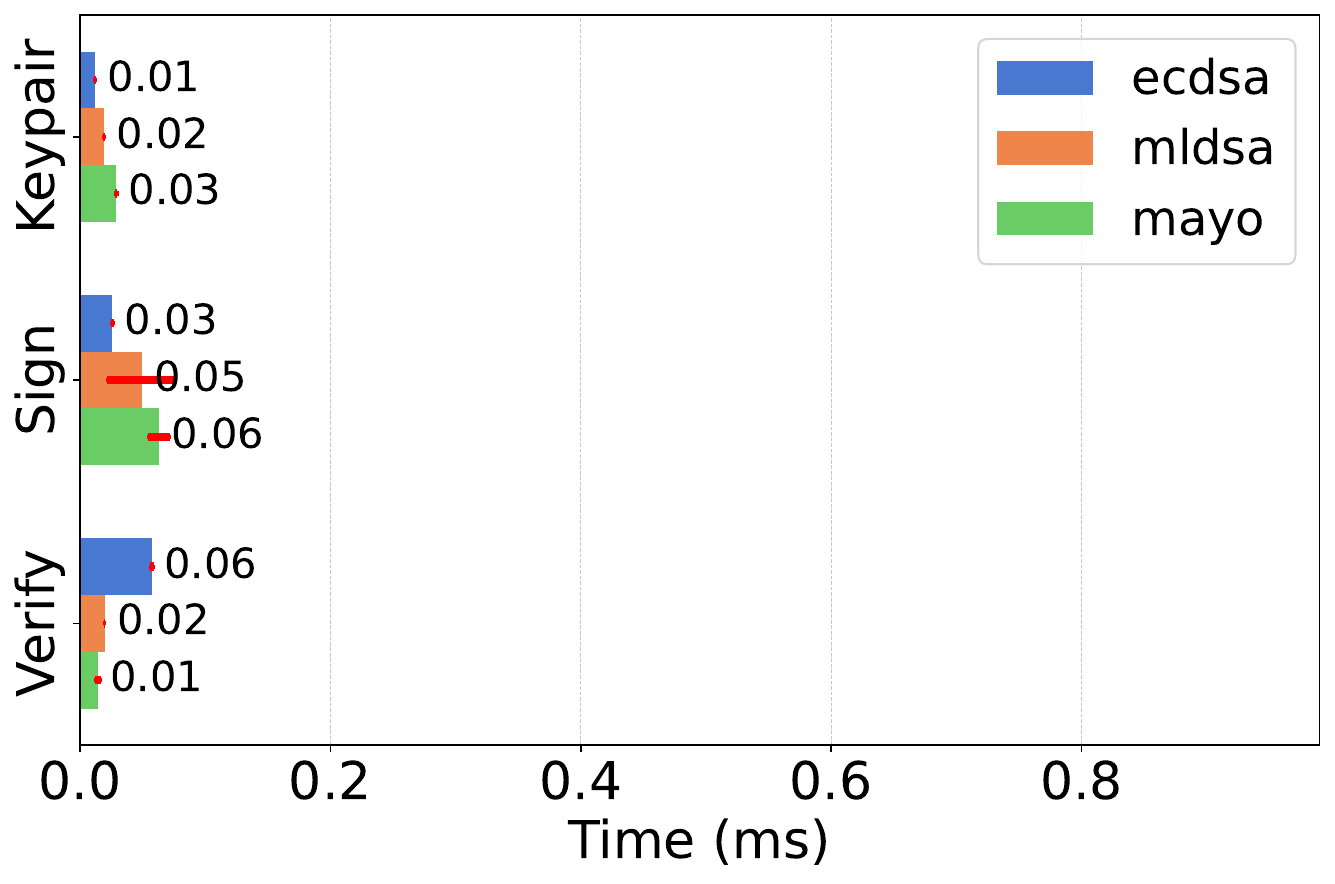}
\label{fig_m2-execucoes-level-1}
}
\subfigure[Laptop x64 - Level 3]{
\includegraphics[width=.31\textwidth]{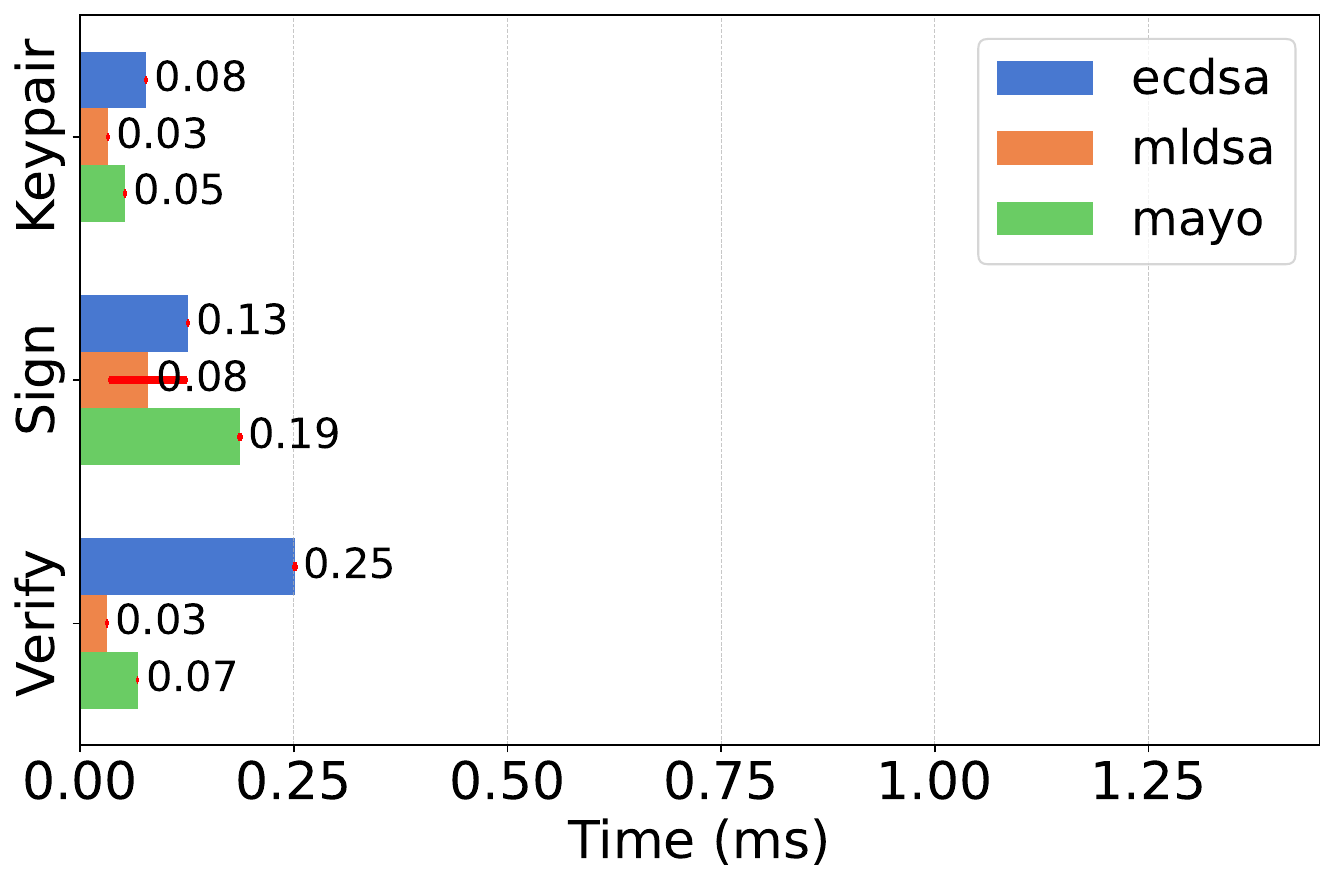}
\label{fig_m2-execucoes-level-3}
}
\subfigure[Laptop x64- Level 5]{
\includegraphics[width=.3\textwidth]{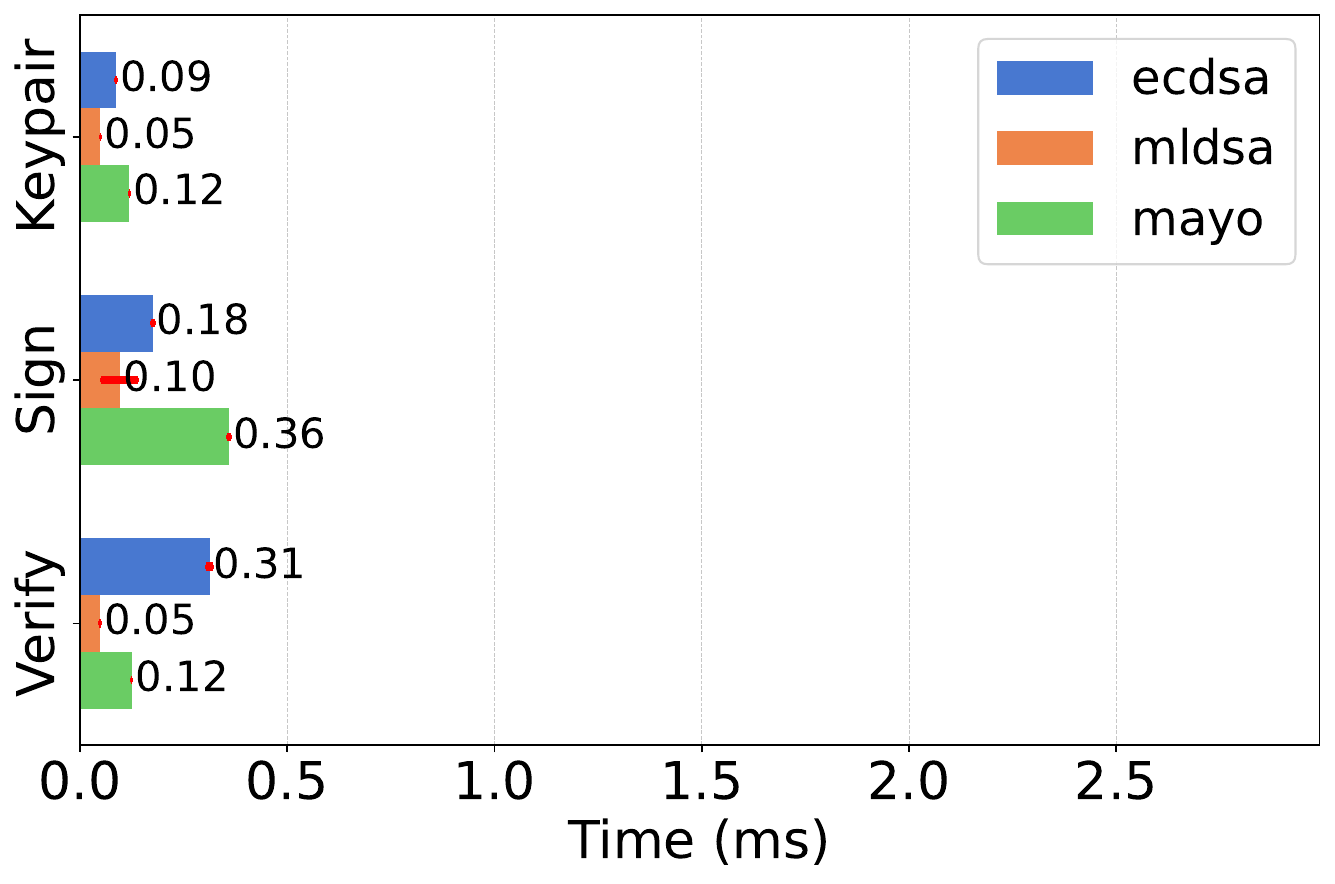}
\label{fig_m2-execucoes-level-5}
}
\subfigure[Desktop - Lower Level (1 or 2)]{
\includegraphics[width=.31\textwidth]{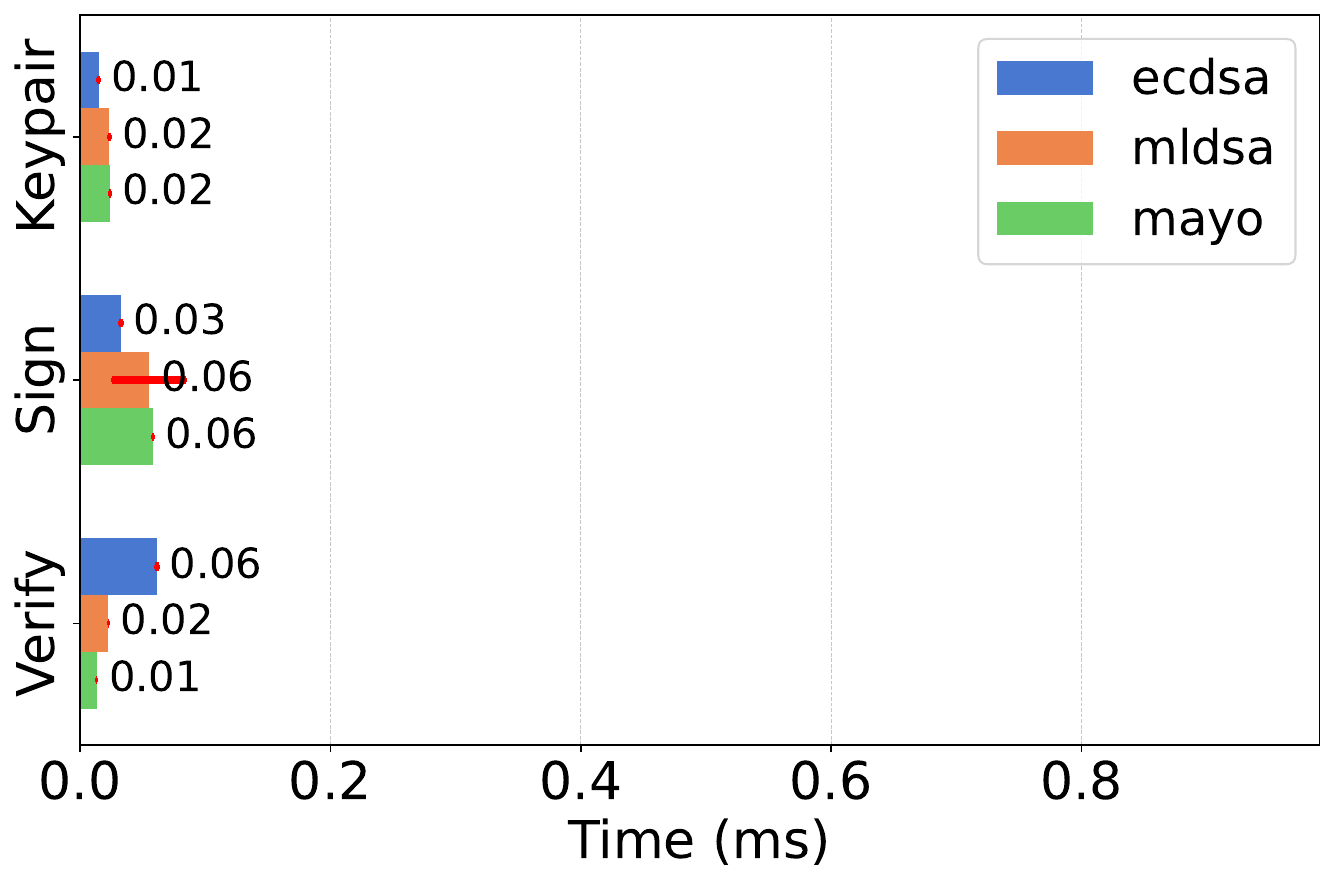}
\label{fig_m3-execucoes-level-1}
}
\subfigure[Desktop - Level 3]{
\includegraphics[width=.31\textwidth]{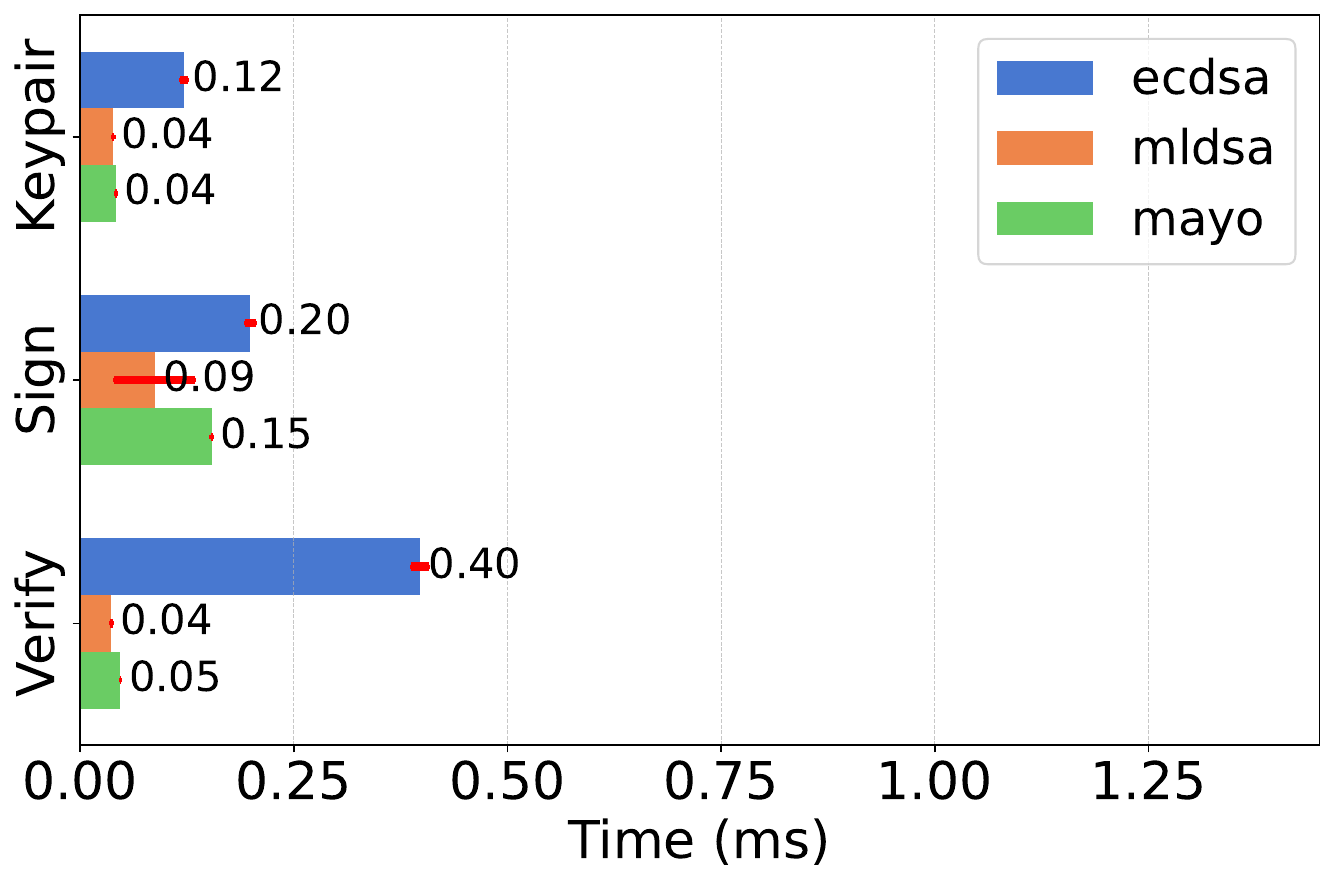}
\label{fig_m3-execucoes-level-3}
}
\subfigure[Desktop - Level 5]{
\includegraphics[width=.31\textwidth]{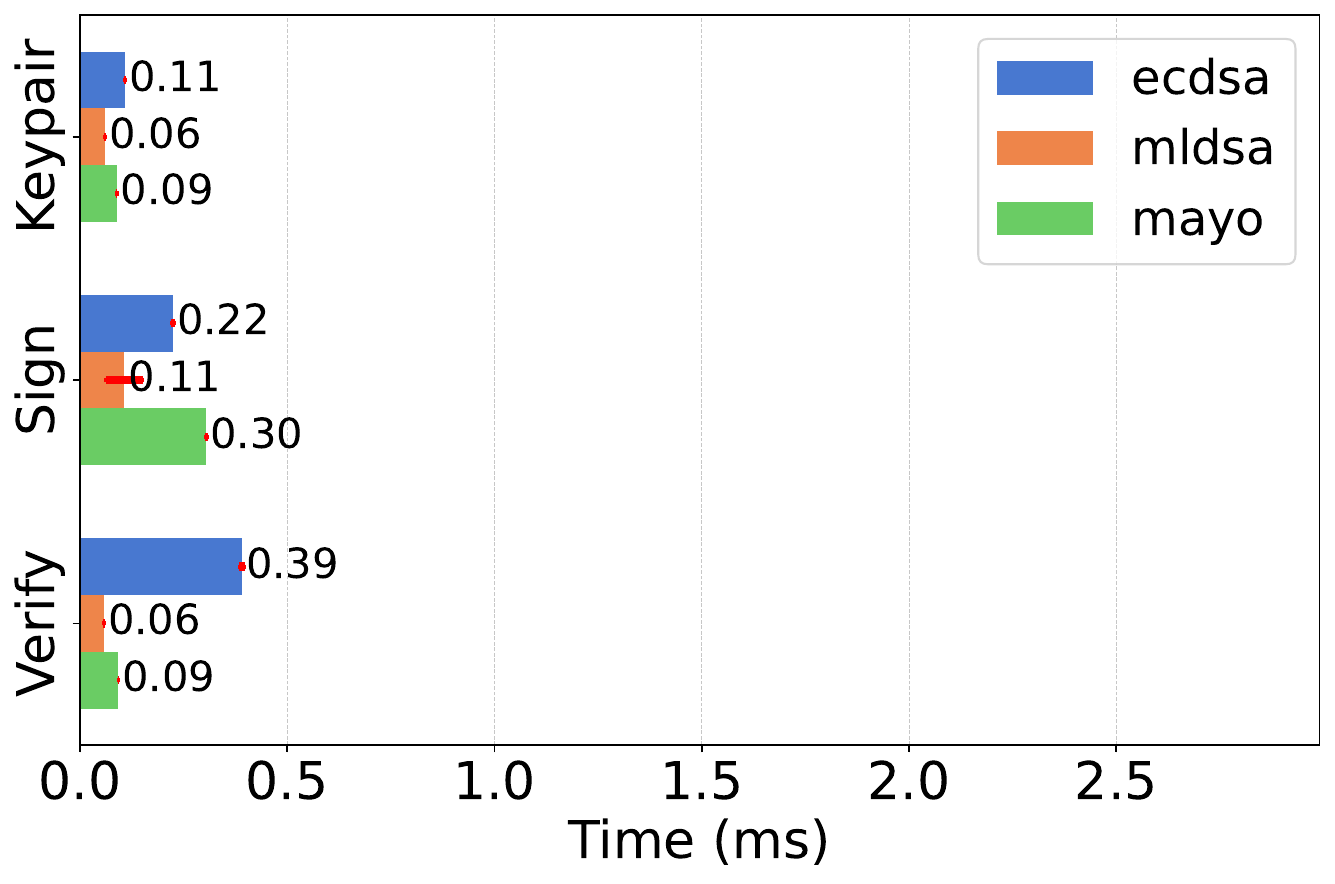}
\label{fig_m3-execucoes-level-5}
}

\caption{Benchmarking: Selected Results (Lower Values Indicate Better Results). }
\label{fig_benchmark}
\end{figure*}

Figure~\ref{fig_benchmark} highlights consistent performance patterns across different hardware configurations and security levels. The results confirm the expected trend of increased computational overhead with higher security levels. Notably, the small standard deviations across all measurements indicate statistical reliability despite potential system-level interference, which validates the adequacy of our sample size.

Our analysis shows that ML-DSA outperforms ECDSA in most operations and security levels, except for Level 1 key-pair generation and signing operations. This comparison is particularly significant since ML-DSA's minimum security level (Level 2) exceeds ECDSA's Level 1, making its performance advantage even more relevant. Mayo exhibits comparable or better performance than ECDSA at Levels 1 and 3 on x64 systems. At Level 5, Mayo demonstrates faster verification but slower key generation and signing operations compared to ECDSA.

The most pronounced performance improvement occurs in verification operations at Level 5 on the ARM-based laptop, where ML-DSA achieves a 92\% reduction in verification time compared to ECDSA (0.06 seconds versus 0.74 seconds). This result demonstrates the potential of post-quantum cryptographic algorithms to maintain or even enhance verification efficiency while providing higher security.

\subsection{Impact on Blockchain Networks}\label{sec:impact-blockchain}

\begin{table*}[!htp]
    \centering
    \scriptsize
    \caption{Complete Results of the blockchain simulations (ms).}
    \label{tab:simulation}
    \renewcommand{\arraystretch}{1.1}
    \setlength{\tabcolsep}{1pt} 
    \resizebox{\textwidth}{!}{
    \begin{tabular}{l||c|c|c||c|c|c||c|c|c}
 
 \hline \hline  
 & \multicolumn{3}{c||}{Laptop ARM} & \multicolumn{3}{c||}{Laptop x64} & \multicolumn{3}{c}{Desktop} \\ 
  & \multicolumn{3}{c||}{verify (mean ± std)} & \multicolumn{3}{c||}{verify (mean ± std)} & \multicolumn{3}{c}{verify (mean ± std)} \\ 

Algorithm  & Lower Level & Level 3 & Level 5 & Lower Level & Level 3 & Level 5 & Lower Level & Level 3 & Level 5  \\ 

\hline \hline \multicolumn{10}{c}{Bitcoin} \\  \hline 

ECDSA & 136.2845 ± 2.3648 & 1273.9468 ± 22.616 & 1531.0027 ± 26.5852 & 99.2422 ± 1.7495 & 434.3056 ± 8.2095 & 541.203 ± 8.8813 & 106.4078 ± 2.0004 & 687.6015 ± 11.907 & 676.0999 ± 12.2533 \\
ML-DSA & 91.3405 ± 1.6492 & 146.6004 ± 2.5224 & 242.2411 ± 4.3684 & 33.8754 ± 0.5687 & 54.3813 ± 0.9583 & 82.9072 ± 1.3698 & 39.1121 ± 0.6682 & 63.6307 ± 1.1611 & 100.0014 ± 1.7721 \\
Dilithium & 38.2409 ± 0.6844 & 56.9484 ± 1.0043 & 90.3939 ± 1.5367 & 33.4214 ± 0.6072 & 54.4299 ± 1.0108 & 85.3526 ± 1.4914 & 38.9487 ± 0.7239 & 63.698 ± 1.1343 & 100.1268 ± 1.8287 \\
Falcon & 50.221 ± 0.8659 &  & 96.4311 ± 1.6821 & 75.0105 ± 1.3633 &  & 104.4389 ± 1.8022 & 58.8385 ± 1.1036 &  & 113.0021 ± 2.0954 \\
Falcon-padded & 49.5613 ± 0.9207 &  & 95.3122 ± 1.7944 & 55.1643 ± 0.9526 &  & 102.4074 ± 1.7481 & 58.4768 ± 1.0218 &  & 111.5594 ± 2.0978 \\
Mayo & 59.1631 ± 1.0231 & 299.4169 ± 5.2516 & 655.7262 ± 12.0629 & 24.5113 ± 0.4112 & 116.3231 ± 2.1785 & 215.494 ± 3.5414 & 22.9609 ± 0.4017 & 81.8565 ± 1.3578 & 159.3144 ± 2.9455 \\
SPHINCS-SHA-s & 528.9003 ± 9.4844 & 871.5926 ± 15.626 & 1260.301 ± 22.5475 & 309.1484 ± 5.5676 & 515.6341 ± 9.228 & 720.2043 ± 12.1136 & 287.0224 ± 5.2349 & 481.1766 ± 8.638 & 662.8648 ± 11.1765 \\
SPHINCS-SHA-f & 1507.3205 ± 25.3508 & 2361.3837 ± 42.4078 & 2434.5585 ± 43.5861 & 853.0385 ± 15.3951 & 1298.8756 ± 22.9854 & 1318.8456 ± 23.079 & 753.3344 ± 14.9031 & 1141.0721 ± 20.085 & 1162.6435 ± 19.3736 \\
SPHINCS-SHAKE-s & 954.4281 ± 16.8238 & 1396.1869 ± 24.1091 & 2065.4039 ± 37.0847 & 599.1115 ± 10.3477 & 847.6319 ± 15.0924 & 1218.7581 ± 20.8864 & 547.3769 ± 9.5828 & 777.7835 ± 13.7176 & 1086.4028 ± 18.9974 \\
SPHINCS-SHAKE-f & 2745.388 ± 45.5068 & 4084.8346 ± 75.6695 & 4180.3499 ± 73.4508 & 1620.9727 ± 30.3867 & 2330.0321 ± 41.1458 & 2366.5064 ± 44.7915 & 1444.0436 ± 25.8178 & 2124.6605 ± 40.8339 & 2074.9598 ± 36.0649 \\
Cross-rsdp-small & 3036.4521 ± 56.3102 & 3987.9731 ± 67.4678 & 5302.8723 ± 101.464 & 2648.0999 ± 47.0922 & 3683.7922 ± 64.0329 & 4917.4178 ± 90.5952 & 2447.8821 ± 46.593 & 3405.9624 ± 58.4809 & 4468.295 ± 81.5143 \\
Cross-rsdpg-small & 2048.0467 ± 34.9935 & 3090.3061 ± 56.5592 & 3897.5645 ± 72.9742 & 1575.7796 ± 27.6206 & 2522.9822 ± 44.7467 & 3083.8917 ± 52.2752 & 1531.6611 ± 25.7086 & 2522.5871 ± 43.5081 & 2900.9881 ± 50.9739 \\
Cross-rsdp-balanced & 797.2513 ± 14.5664 & 1732.6422 ± 33.8118 & 2947.9559 ± 53.5516 & 675.7915 ± 11.7194 & 1508.6183 ± 27.3333 & 2489.3299 ± 47.2395 & 636.0842 ± 11.019 & 1427.7683 ± 24.8086 & 2331.9637 ± 43.0686 \\
Cross-rsdpg-balanced & 582.2241 ± 10.2936 & 817.1568 ± 14.0075 & 1409.3277 ± 24.231 & 449.4047 ± 7.7085 & 693.8905 ± 12.7111 & 1137.5636 ± 21.8478 & 428.8288 ± 7.5287 & 666.3342 ± 11.1477 & 1068.4216 ± 18.7796 \\
Cross-rsdp-fast & 450.1529 ± 9.1348 & 1047.6817 ± 16.8568 & 1997.015 ± 35.2195 & 363.3275 ± 6.6688 & 841.1026 ± 14.1928 & 1452.9939 ± 26.2099 & 340.7624 ± 5.5979 & 788.0883 ± 14.5187 & 1373.4859 ± 23.8799 \\
Cross-rsdpg-fast & 324.5005 ± 5.9552 & 647.3387 ± 11.9267 & 1154.7252 ± 19.8287 & 252.0224 ± 4.701 & 596.4249 ± 10.2094 & 973.0883 ± 18.1496 & 234.9852 ± 4.4663 & 561.3991 ± 10.1477 & 897.5819 ± 15.9627 \\

\hline \hline \multicolumn{10}{c}{Ethereum} \\  \hline 

ECDSA & 10.3491 ± 0.2384 & 96.6628 ± 2.1812 & 116.0763 ± 2.6017 & 7.4872 ± 0.1458 & 32.7918 ± 0.6453 & 40.8184 ± 0.7706 & 8.1537 ± 0.1449 & 52.7183 ± 0.9651 & 51.8054 ± 0.9224 \\
ML-DSA & 6.9448 ± 0.1578 & 11.1379 ± 0.2465 & 18.418 ± 0.4154 & 2.5527 ± 0.0491 & 4.1037 ± 0.0793 & 6.2515 ± 0.1186 & 2.9991 ± 0.0547 & 4.8817 ± 0.0907 & 7.6552 ± 0.1401 \\
Dilithium & 2.9024 ± 0.0643 & 4.3213 ± 0.0929 & 6.8603 ± 0.1544 & 2.5226 ± 0.0514 & 4.1034 ± 0.0833 & 6.427 ± 0.1325 & 2.9824 ± 0.0535 & 4.8851 ± 0.0829 & 7.6745 ± 0.1331 \\
Falcon & 3.8116 ± 0.0863 &  & 7.3194 ± 0.1698 & 5.6556 ± 0.1117 &  & 7.8667 ± 0.1552 & 4.5113 ± 0.0813 &  & 8.6582 ± 0.1592 \\
Falcon-padded & 3.7618 ± 0.0863 &  & 7.2435 ± 0.1603 & 4.1596 ± 0.0772 &  & 7.7213 ± 0.1537 & 4.4858 ± 0.0811 &  & 8.5432 ± 0.1531 \\
Mayo & 4.4919 ± 0.1002 & 22.7227 ± 0.5212 & 49.7921 ± 1.1234 & 1.8454 ± 0.0368 & 8.7817 ± 0.1695 & 16.2426 ± 0.3261 & 1.7588 ± 0.0314 & 6.2742 ± 0.1108 & 12.2031 ± 0.222 \\
SPHINCS-SHA-s & 40.173 ± 0.903 & 66.1663 ± 1.5228 & 95.5925 ± 2.233 & 23.3252 ± 0.4345 & 38.8725 ± 0.781 & 54.25 ± 1.0775 & 22.0238 ± 0.3913 & 36.831 ± 0.67 & 50.7323 ± 0.8998 \\
SPHINCS-SHA-f & 114.4394 ± 2.573 & 179.4633 ± 4.022 & 184.6904 ± 4.1013 & 64.2922 ± 1.2847 & 97.9234 ± 1.9218 & 99.2531 ± 1.9502 & 57.7019 ± 1.0366 & 87.5299 ± 1.5782 & 89.0286 ± 1.5921 \\
SPHINCS-SHAKE-s & 72.6221 ± 1.6532 & 105.8964 ± 2.4491 & 156.9101 ± 3.5692 & 45.1136 ± 0.9134 & 63.956 ± 1.2467 & 91.8136 ± 1.7776 & 41.896 ± 0.7654 & 59.6529 ± 1.0579 & 83.1413 ± 1.4885 \\
SPHINCS-SHAKE-f & 208.2749 ± 4.8916 & 310.2855 ± 6.8458 & 317.4644 ± 7.2003 & 122.203 ± 2.3795 & 175.5756 ± 3.4583 & 178.4881 ± 3.3593 & 110.7192 ± 2.0648 & 162.7238 ± 3.0224 & 158.9173 ± 2.8266 \\
Cross-rsdp-small & 230.9886 ± 4.9499 & 302.7863 ± 7.0252 & 403.1348 ± 9.0603 & 199.7076 ± 4.0614 & 277.5764 ± 5.2626 & 370.2685 ± 7.317 & 187.608 ± 3.4582 & 260.9108 ± 4.6804 & 342.4696 ± 6.2811 \\
Cross-rsdpg-small & 155.3068 ± 3.6172 & 234.7216 ± 5.221 & 295.9017 ± 7.0679 & 118.6946 ± 2.3037 & 190.0899 ± 3.675 & 232.5503 ± 4.6263 & 117.2321 ± 2.1072 & 193.0565 ± 3.5995 & 222.1975 ± 4.0865 \\
Cross-rsdp-balanced & 60.5409 ± 1.3722 & 131.6209 ± 2.9952 & 224.051 ± 4.8656 & 50.9738 ± 0.978 & 113.8022 ± 2.2499 & 188.1103 ± 3.7592 & 48.732 ± 0.9241 & 109.3643 ± 1.8698 & 178.8018 ± 3.2693 \\
Cross-rsdpg-balanced & 44.2783 ± 0.9875 & 62.0173 ± 1.4034 & 107.04 ± 2.3375 & 33.937 ± 0.6676 & 52.3021 ± 1.0433 & 85.7211 ± 1.6835 & 32.8924 ± 0.5854 & 51.0591 ± 0.8984 & 81.8082 ± 1.5069 \\
Cross-rsdp-fast & 34.1856 ± 0.7596 & 79.5323 ± 1.7342 & 151.6858 ± 3.2924 & 27.3821 ± 0.5532 & 63.4258 ± 1.2638 & 109.5058 ± 2.065 & 26.0865 ± 0.471 & 60.2854 ± 1.1056 & 105.2215 ± 1.9187 \\
Cross-rsdpg-fast & 24.6787 ± 0.5415 & 49.2222 ± 1.0755 & 87.569 ± 2.0809 & 19.0112 ± 0.3663 & 44.8927 ± 0.9238 & 73.4431 ± 1.4179 & 18.0222 ± 0.3291 & 43.0699 ± 0.7775 & 68.8634 ± 1.2179 \\

\hline \hline 
\end{tabular}}
\end{table*}

Table~\ref{tab:simulation} presents the complete blockchain simulation results, including mean execution times and standard deviations for block verification operations across all evaluated security levels in Bitcoin and Ethereum networks. 

\begin{figure*}[h!]
\centering
\subfigure[Laptop ARM - Lower Level (1 or 2)]{
\includegraphics[width=.31\textwidth]{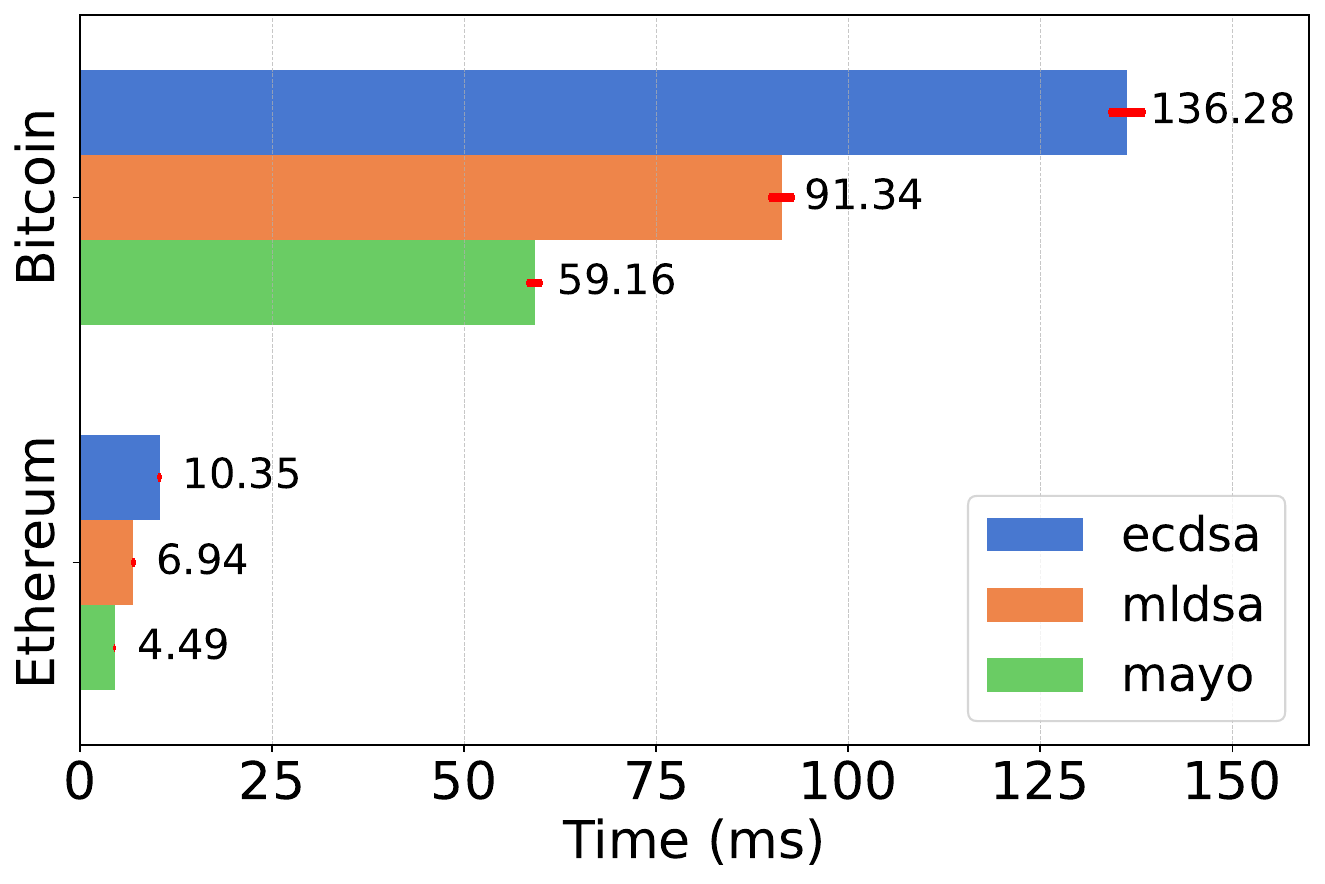}
\label{fig_m1-blocksim-level-1}
}
\subfigure[Laptop ARM - Level 3]{
\includegraphics[width=.31\textwidth]{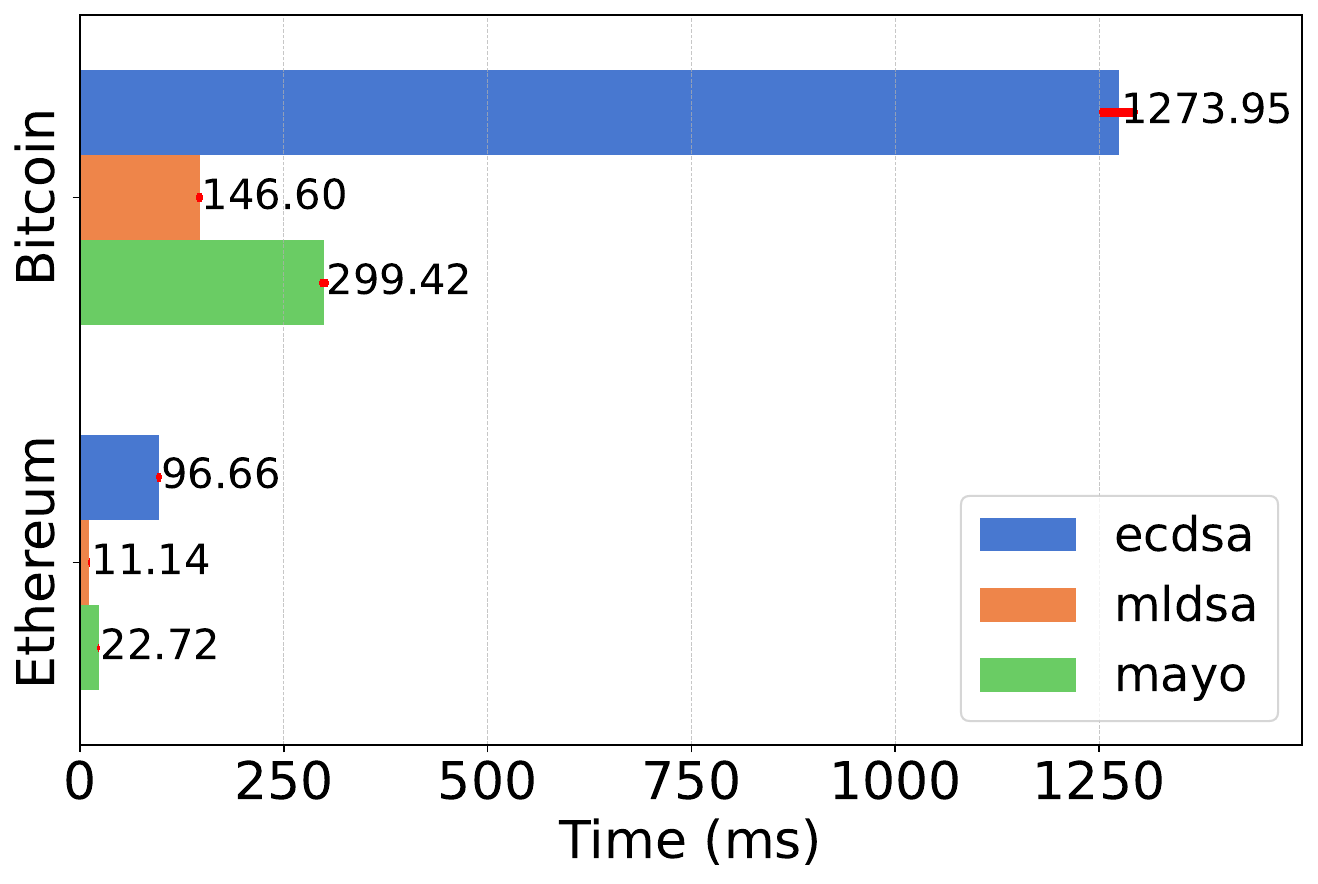}
\label{fig_m1-blocksim-level-3}
}
\subfigure[Laptop ARM - Level 5]{
\includegraphics[width=.31\textwidth]{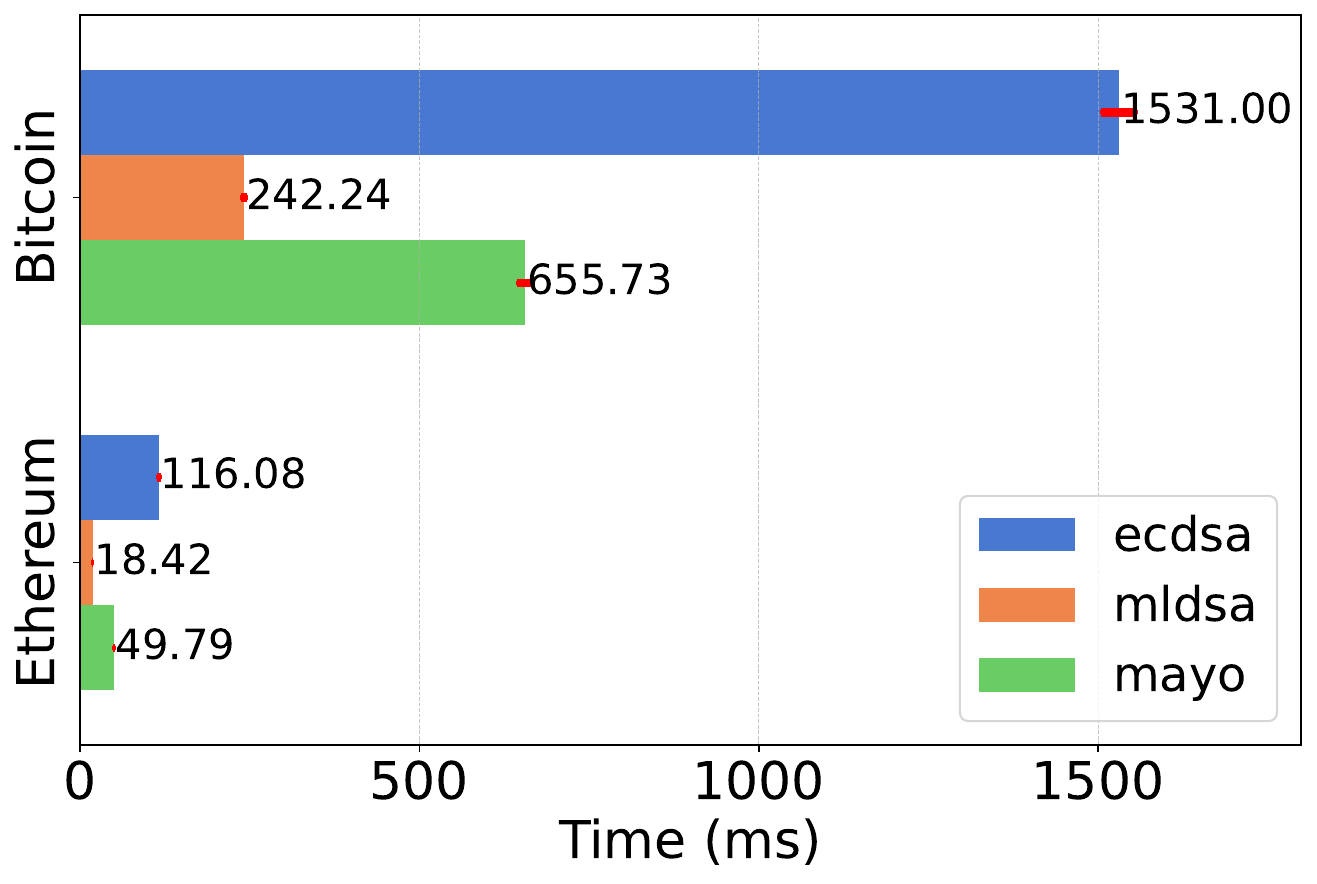}
\label{fig_m1-blocksim-level-5}
}
\subfigure[Laptop x64 - Lower Level (1 or 2)]{
\includegraphics[width=.31\textwidth]{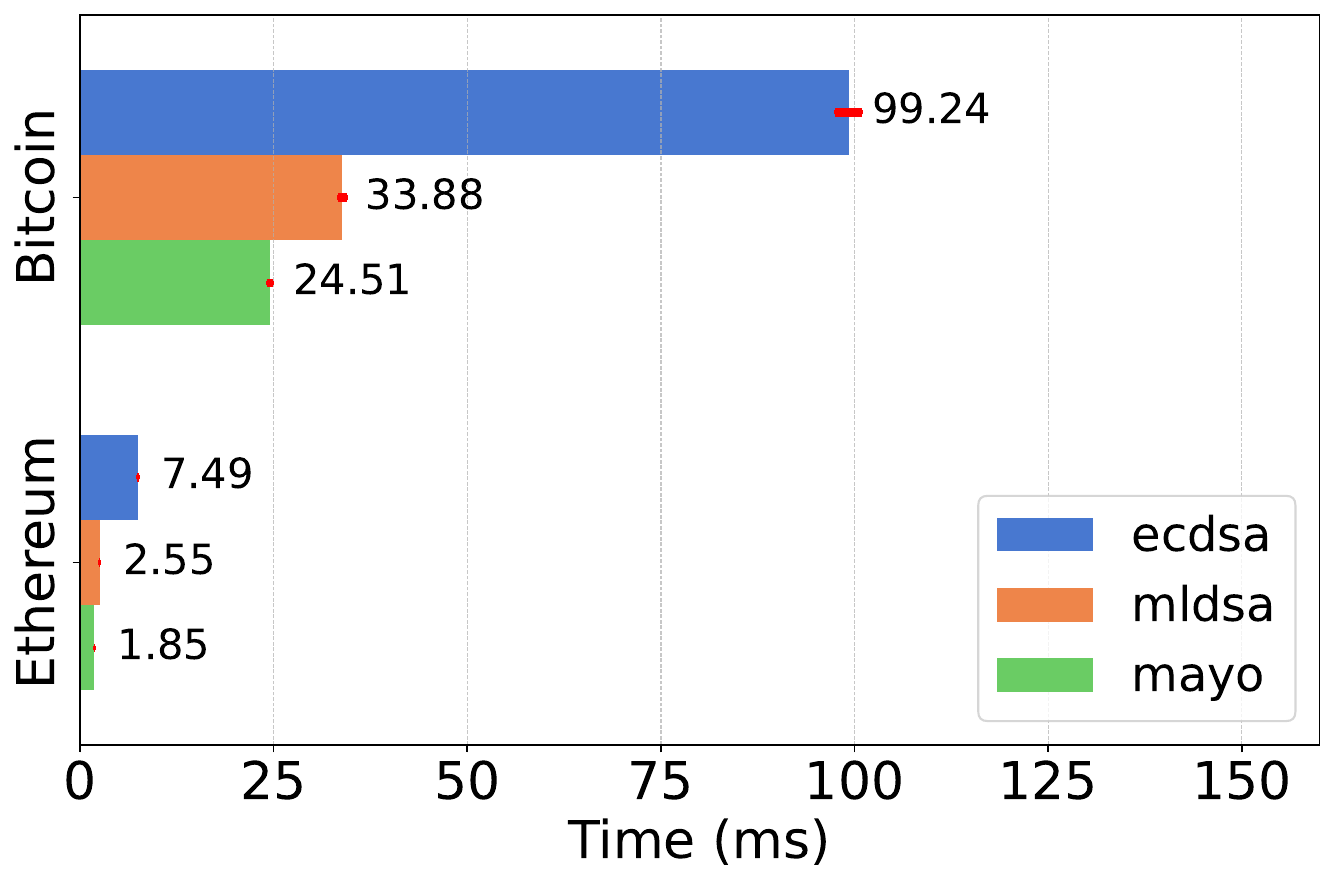}
\label{fig_m2-blocksim-level-1}
}
\subfigure[Laptop x64  - Level 3]{
\includegraphics[width=.31\textwidth]{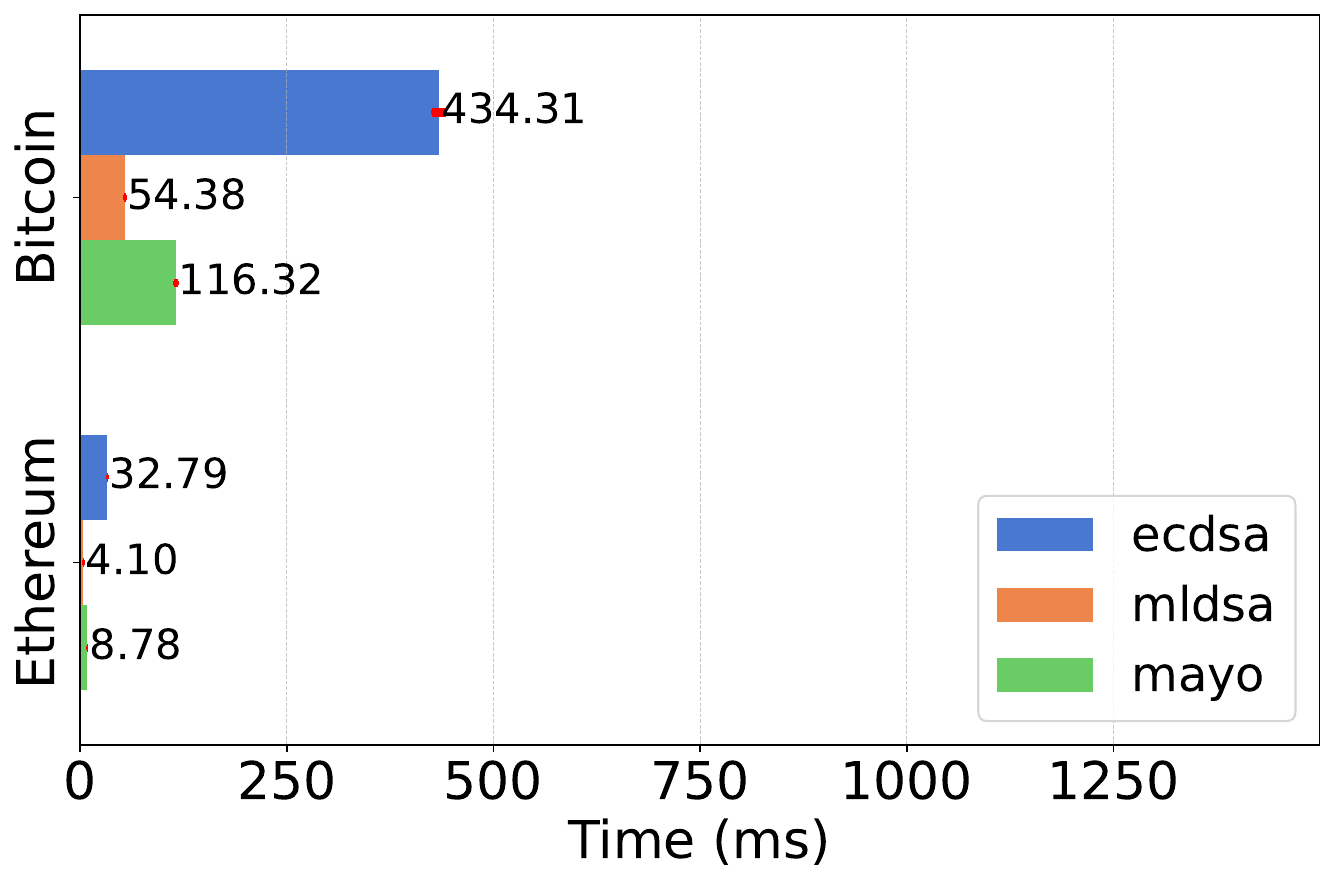}
\label{fig_m2-blocksim-level-3}
}
\subfigure[Laptop x64  - Level 5]{
\includegraphics[width=.31\textwidth]{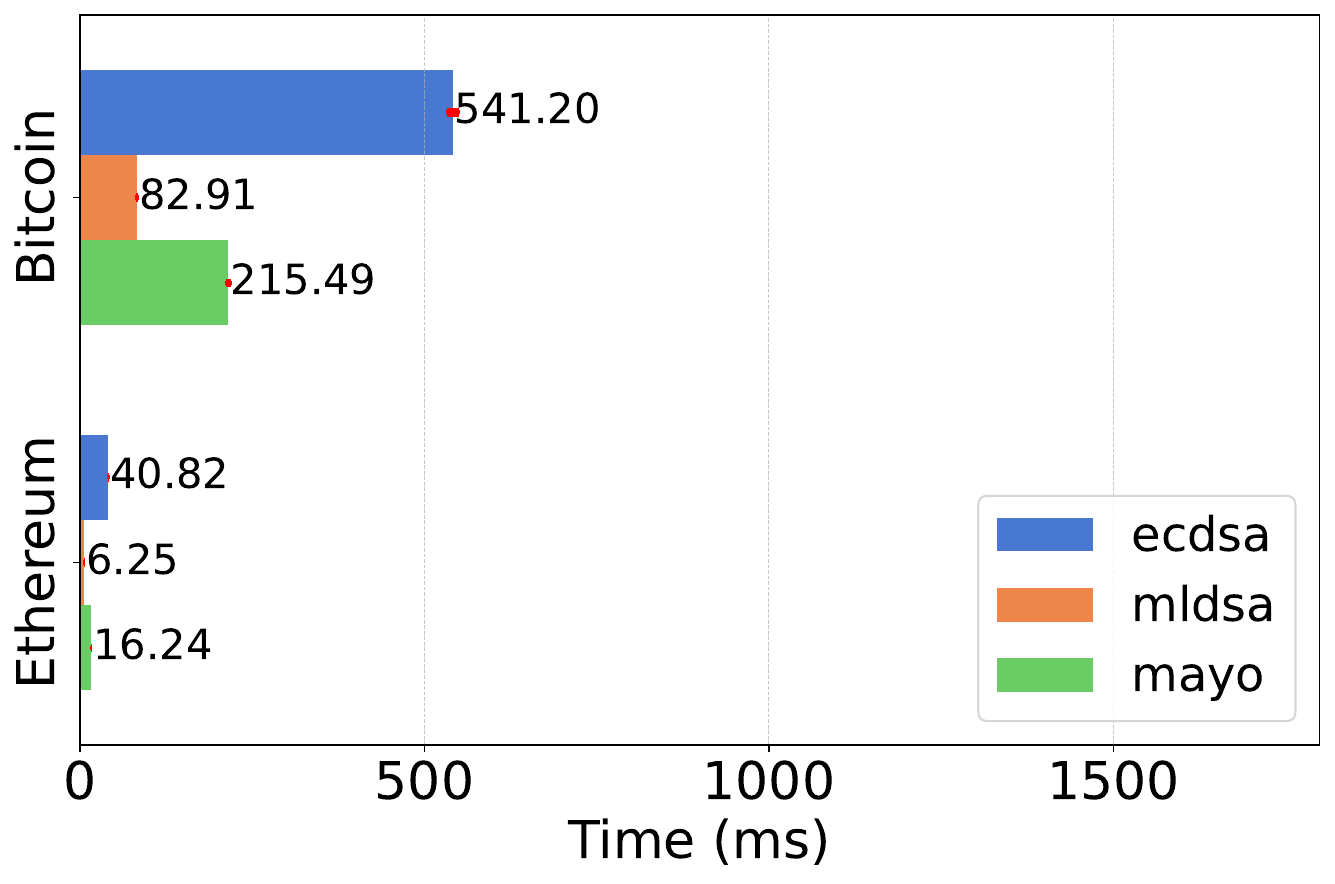}
\label{fig_m2-blocksim-level-5}
}
\subfigure[Desktop - Lower Level (1 or 2)]{
\includegraphics[width=.31\textwidth]{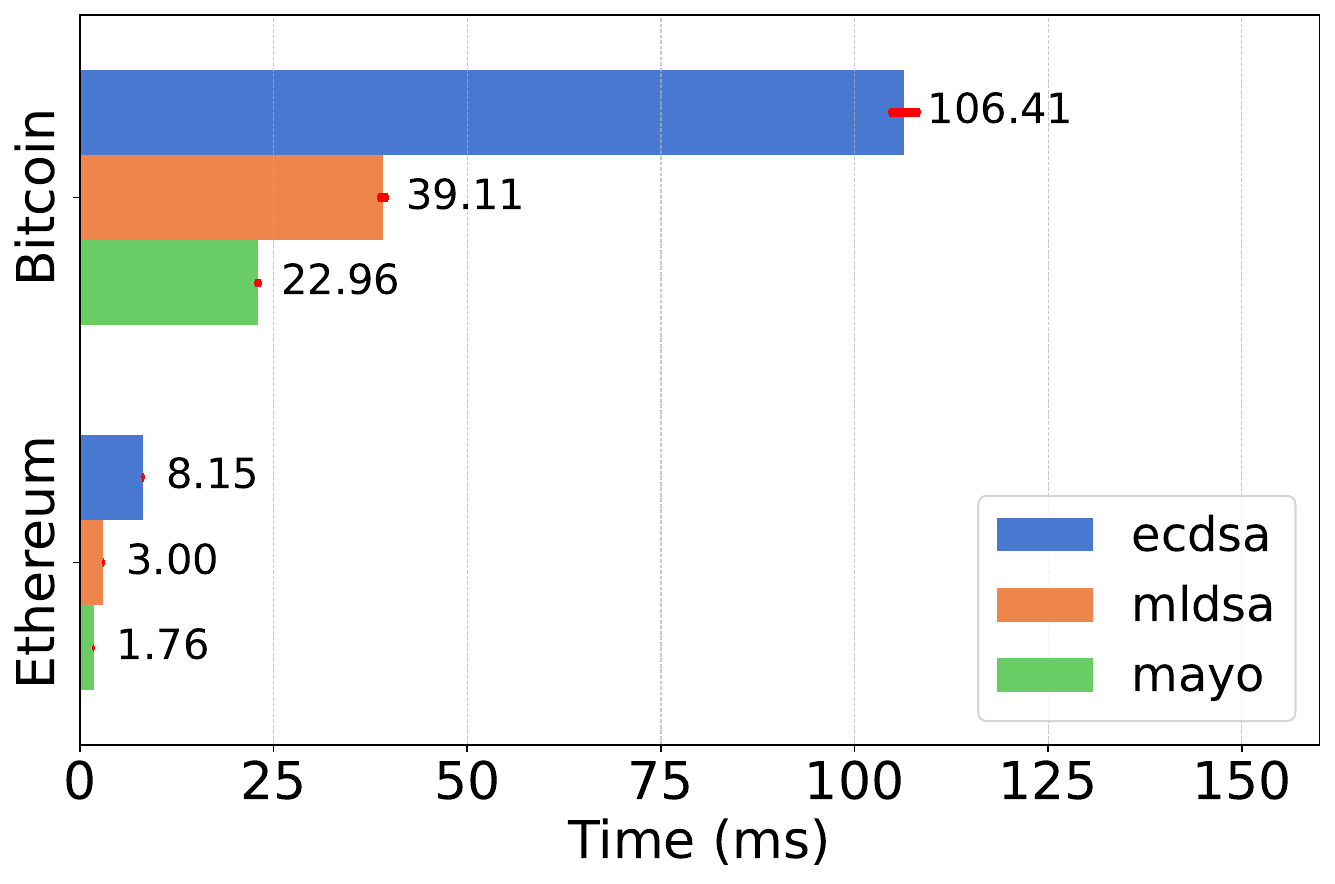}
\label{fig_m3-blocksim-level-1}
}
\subfigure[Desktop - Level 3]{
\includegraphics[width=.31\textwidth]{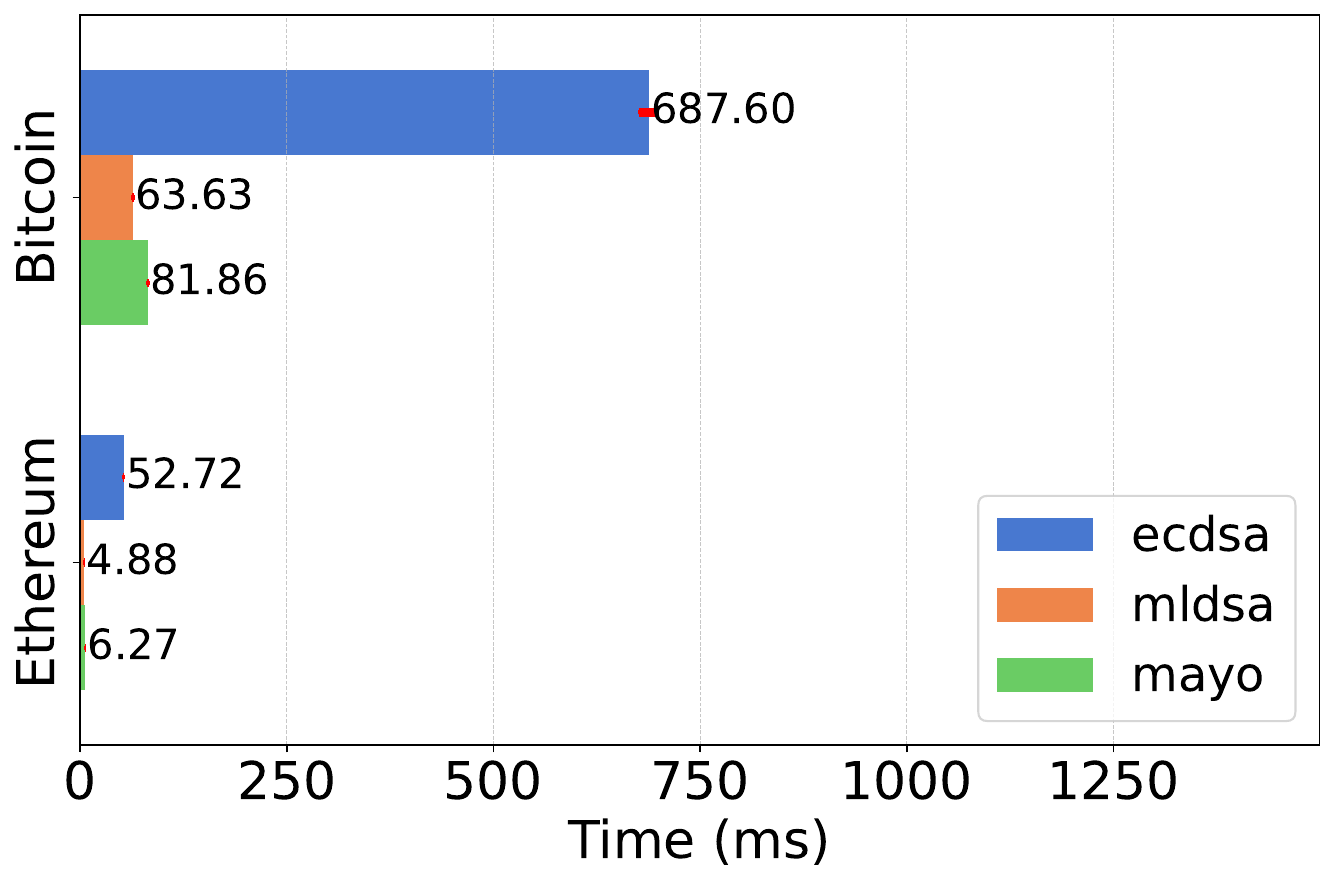}
\label{fig_m3-blocksim-level-3}
}
\subfigure[Desktop - Level 5]{
\includegraphics[width=.31\textwidth]{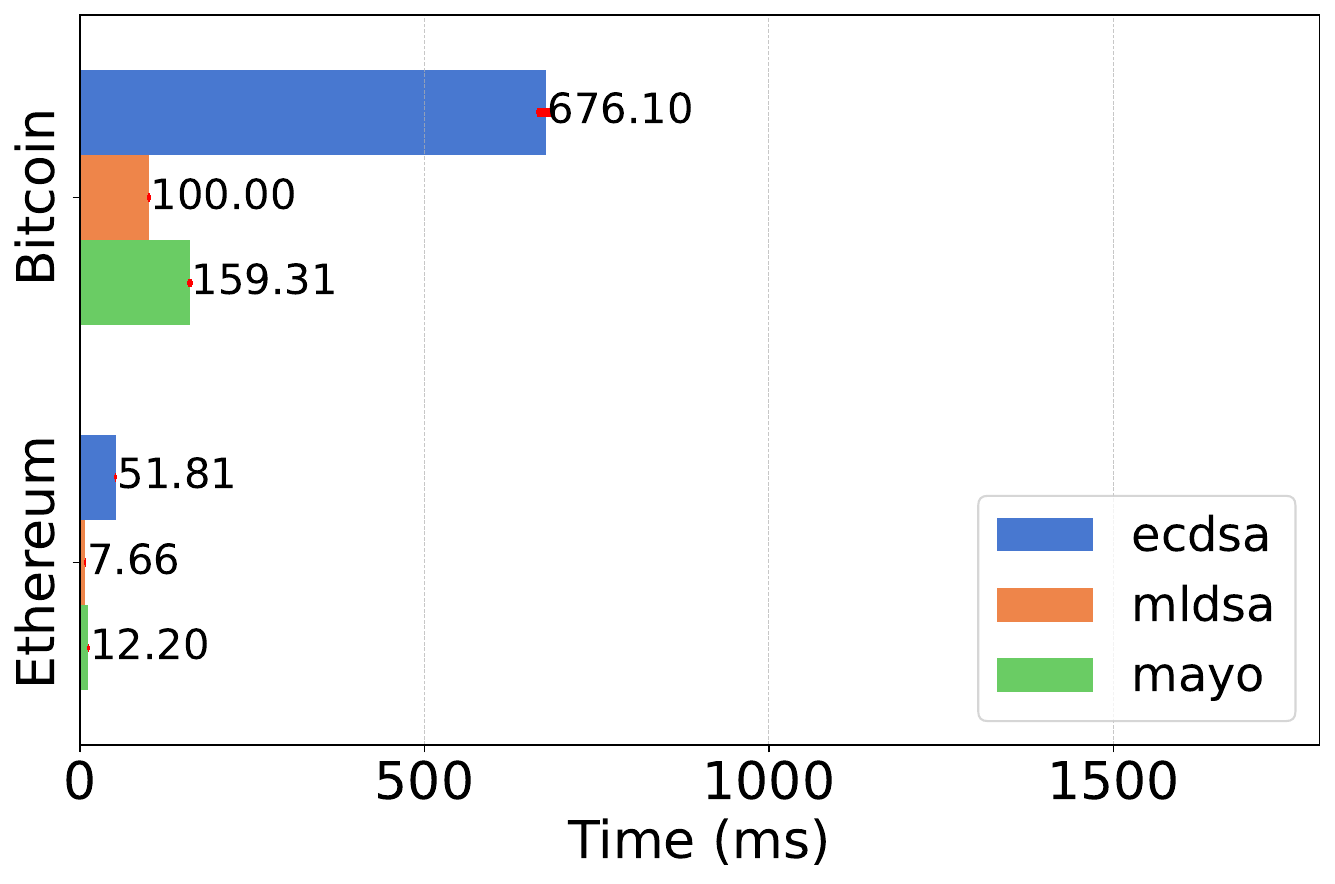}
\label{fig_m3-blocksim-level-5}
}
\caption{Main Simulation Results (Lower Values Indicate Better Results).}
\label{fig_blocksim-results}
\end{figure*}

Figure~\ref{fig_blocksim-results} presents the performance of the two most efficient post-quantum candidates, ML-DSA and Mayo, alongside the traditional algorithm ECDSA. The results show that in blockchain network simulations, post-quantum cryptographic algorithms consistently outperform ECDSA across all evaluated scenarios. Both ML-DSA and Mayo achieve superior performance, with particularly notable gains at higher security levels. In particular, ML-DSA reduces execution time by 90\% at Level 3 (63.63 ms compared to 687.60 ms for ECDSA) on a Desktop system, demonstrating the feasibility of adopting PQC algorithms without compromising system performance.

\subsection{Discussion}

As the experiments demonstrate, some PQC algorithms perform similarly to, and in some cases outperform, ECDSA, both individually and when applied in blockchain systems. Therefore, migrating digital signatures from ECDSA to PQC algorithms, particularly ML-DSA and Mayo, can be performed smoothly in systems where application performance is the primary requirement. However, migrating to other algorithms, such as SLH-DSA and Cross-rsdp, is not recommended in these scenarios due to performance limitations.

Although this study did not evaluate the sizes of public keys, private keys, and digital signatures, previous works \cite{benchmarkPQCPi4,classicalVsPQCinPi4-desktop-laptop,PQCinEthereum-based} and published standards \cite{fips-204, fips-205} indicate that PQC-generated keys and signatures are considerably larger than those of ECDSA. Except SLH-DSA, which has the same key size as ECDSA. Systems with storage constraints, including blockchain networks and IoT devices, must balance application security requirements against the capacity to store these larger keys and signatures when deciding which algorithms to adopt.

Moreover, adopting PQC algorithms in blockchains requires considering not only computational performance but also the impact of larger keys and signatures on transaction and block sizes. Larger signatures increase transaction size, reducing the number of transactions per block. Consequently, larger blocks with fewer transactions can elevate latency for consensus or transaction confirmation, potentially affecting overall network throughput.

\subsection{Limitations}

Comparisons with other traditional algorithms, such as RSA, were not conducted in this study. Additionally, discussions on the adoption and potential improvements of key encapsulation mechanisms (KEM) \cite{trustcom1} are beyond the scope of this work. Our focus is specifically on processing involving digital signatures, as employed in the most widely used blockchains, including Bitcoin and Ethereum.

Although some proposed solutions explore blockchain applications in IoT contexts \cite{Dorri:2017, Lunardi:2022, trustcom2}, we chose to concentrate our experiments on the most prevalent blockchain systems. Furthermore, \texttt{BlockSim} currently supports only Bitcoin and Ethereum, which are not commonly used in IoT deployments. Evaluating the impact of PQC algorithms in IoT devices remains an avenue for future research.

\section{Conclusion}\label{sec:considerations}

In this work, we presented a comparison between the performance of traditional algorithms, such as ECDSA, and post-quantum cryptographic (PQC) algorithms, including ML-DSA, Falcon, and SPHINCS+, both in isolation and within blockchain systems. We also introduced a tool that benchmarks the core operations of these algorithms—key generation, signing, and verification—and simulates their impact on blockchain.

Our results indicate that certain PQC algorithms, particularly ML-DSA and Mayo, outperform ECDSA, especially at higher security levels. As expected, performance decreases as security levels increase, reflecting the corresponding rise in computational overhead. These findings suggest that replacing algorithms vulnerable to Shor's algorithm is feasible, provided the effects of larger key and signature sizes on network throughput are considered.

As Future work, we intend to investigate the impact of public key, private key and digital signature sizes on blockchain storage and network traffic, evaluating their implications for system efficiency, scalability, and operational viability. Moreover, the modular architecture of \texttt{PQCinBlock} allows the integration of new algorithms, programming languages, and experimental scenarios, enabling the tool to support both traditional and post-quantum digital signature schemes in future studies.

\bibliographystyle{IEEEtran}
\bibliography{references}

\end{document}